\documentclass[aps,tightenlines,twocolumn,floats,prd,nofootinbib,superscriptaddress,showpacs
]{revtex4-1}
\def\bea{\begin{eqnarray}}
\def\eea{\end{eqnarray}}
\def\bal{\begin{align}}
\def\eal{\end{align}}

\def\sfrac#1#2{{\textstyle \frac{#1}{#2}}}

\usepackage[dvips]{color}
\usepackage{graphics}
\usepackage{graphicx}
\usepackage{epsf} 
\usepackage{amsmath}
\usepackage{amssymb}
\usepackage{bbold}
\usepackage{bm}
\usepackage{slashed}
\usepackage{float}
\usepackage{psfrag}
\usepackage{footnote} 
\usepackage{xcolor}

\begin{document}

\hspace{5in}\parbox{1.5in}{ \leftline{JLAB-THY-13-1814, CFTP/13-026}
                \leftline{}\leftline{}\leftline{}\leftline{}
}

\title
{\bf Confinement, quark mass functions, and spontaneous chiral symmetry breaking\\ in Minkowski space
}

\author{Elmar P. Biernat}
\affiliation{Centro de F\'isica Te\'orica de Part\'iculas (CFTP), 
Instituto Superior T\'ecnico (IST), Universidade de Lisboa, 
Av. Rovisco Pais, 1049-001 Lisboa, Portugal}
\author{Franz Gross }
 \affiliation{ Thomas Jefferson National Accelerator Facility (JLab), Newport News, VA 23606, USA}
\author{M.~T.~Pe\~na}
\affiliation{Centro de F\'isica Te\'orica de Part\'iculas (CFTP), 
Instituto Superior T\'ecnico (IST), Universidade de Lisboa, 
Av. Rovisco Pais, 1049-001 Lisboa, Portugal}
\author{Alfred Stadler}
  \affiliation{ Departamento de F\'isica da Universidade de \'Evora, 7000-671 \'Evora, Portugal \\and Centro de F\'isica Nuclear da Universidade de Lisboa (CFNUL), 1649-003 Lisboa, Portugal}
\date{\today}
\begin{abstract}

We formulate the covariant equations for quark-antiquark bound states in Minkowski space in the framework of the Covariant Spectator Theory. The quark propagators are dressed with the same kernel that describes the interaction between different quarks. We show that these equations are charge conjugation invariant, and that in the chiral limit of vanishing bare quark mass, a massless pseudoscalar bound state is produced in a Nambu--Jona-Lasinio (NJL) mechanism, which is associated with the Goldstone boson of spontaneous chiral symmetry breaking. 
In this introductory paper we test the formalism by using a simplified kernel consisting of a momentum-space $\delta$-function with a vector Lorentz structure, to which one adds a mixed scalar and vector confining interaction. The scalar part of the confining interaction is not chirally invariant by itself, but decouples from the equations in the chiral limit and therefore allows the NJL mechanism to work. With this model we calculate the quark mass function, and we compare our Minkowski-space results to LQCD data obtained in Euclidean space.  In a companion paper we apply this formalism to a calculation of the pion form factor.

\end{abstract}

\phantom{0}

\maketitle

\section{Introduction}

The connection between the mass spectrum and structure of hadrons and their underlying internal gluon-quark dynamics is today an important and challenging problem in physics. At the heart of  forthcoming experimental programs at for instance JLab and FAIR, and of the immense progress in lattice QCD (LQCD) calculations \cite{Edwards,Guo} lies the identification of hybrid baryons, exotic and hybrid mesons, as well as the understanding of the full implications of dynamical chiral symmetry breaking. At present, models are necessary to establish a link between LQCD calculations and experimental data.

Furthermore, today a believable model must go beyond the extensive work of Isgur and Godfrey \cite{Godfrey} and Spence and Vary \cite{Spence}. These models might be adequate for heavy-quark systems well described within a nonrelativistic framework, but mesons containing at least one light quark require a fully relativistic treatment. It is therefore desirable for a unified description of all quark-antiquark states to use a bound-state equation that is both Lorentz covariant and that reduces to the Schr\"{o}dinger equation in the non-relativistic limit. In addition, these earlier quark models ignored chiral symmetry breaking and used static potentials (often  variations of the very successful Cornell potential~\cite{Eichten:1975,Eichten:1978,Richardson:1978bt}) to describe the interaction between constituent quarks with a fixed mass. 

Non-perturbative methods, including numerical solutions of QCD on a discrete space-time lattice and of the modern Dyson-Schwinger and mass gap equation \cite{Bars:1977ud,Amer:1983qa,LeYaouanc:1983it,LeYaouanc:1983iy,LeYaouanc:1984dr,Bicudo:1989sh,Bicudo:1989si,Bicudo:1989sj,Bicudo:1993yh,Bicudo:1998mc,Nefediev:2004by,Roberts:2000aa,Alkofer:2000wg,Maris:2003vk,Fischer:2006ub,Rojas:2013tza}, have developed  tools which unify the explanation of a wide range of meson and baryon phenomena \cite{Wilson,SanchisAlepuz:2011jn,Eichmann:2012mp}.  A key feature of these approaches is the emergence of a dynamical description of a constituent quark which can acquire, through dynamical chiral symmetry breaking, a momentum dependent mass function leading to an effective constituent  quark mass which can be much larger than the current-quark mass. It is now possible to discuss the structure of the constituent quark, or how, in the chiral limit, the spontaneous generation of a constituent quark mass is linked to 
the existence of the pion bound state with zero mass through the famous NJL mechanism.  This connection is both interesting and an essential ingredient to our 
understanding of 
the light meson spectrum.  Without it, the light pion mass appears to be an accident resulting from fine-tuning of the interaction parameters.

A covariant treatment of mesons including the NJL-mechanism can be found in the papers by C.\ D.\ Roberts and his collaborators, (see for example \cite{Qin2011} and \cite{Qin2012} and references therein), based on the Dyson-Schwinger equation, which we denote by DSE.  The work presented here starts from a physical model of dynamical quarks and mesons similar to that assumed by the DSE, but it differs in two important aspects: (i) we work in Minkowski space instead of Euclidean space, and (ii) we include a Lorentz scalar confining interaction.   Both of these differences are significant and have their own advantages.

 Aspect (i) is important because, while some observables can be calculated just as well in Euclidean space as in Minkowski space (particle masses, for example), others, such as transition form factors in the timelike region, require a framework defined in Minkowski space.  
Aspect (ii) enables us to address an important question: although phenomenological approaches and lattice calculations may suggest that the confining force is scalar, those indications are not definitive \cite{Kalashnikova:2005tr,Bicudo:2003ji,Allen:2000sd,Michael:1985rh,Gromes:1984ma,Bali:1997am,Lucha:1991vn}, and while DSE suggests that confinement is not important to (and maybe even absent from) the light meson spectrum, it is  clearly an advantage to have an approach that is flexible enough to allow the confining interaction to be present (and to investigate the relative strengths of scalar and vector components).   This is certainly possible in the approach that is developed in this paper.  As it turns out, the scalar confining interaction, when its relativistic extension is adequately defined, decouples in the chiral limit, allowing the NJL mechanism to work. Therefore the breaking of chiral symmetry by such a scalar component does not constitute a problem in our formulation. This remarkable fact will 
be discussed in some detail in this paper.

This work extends and improves upon early work done by Gross and Milana \cite{Gross:1991te,Gross:1991pk,Gross:1994he}, and \c{S}avkli and Gross \cite{Savkli:1999me}, denoted collectively by GMS.  The Covariant Spectator Theory (CST) \cite{Gro69,Gro74,Gro82} already used in GMS, was  extensively tested in the treatment of nucleon-nucleon scattering \cite{Gro08,Gross:2010qm} and pion nucleon scattering \cite{Gro93,Gro96}.  This paper improves on GMS and prepares the way for subsequent studies of the structure of all mesons,  planned for future papers. The formulation shown in this paper differs from previous CST models in the sense that the mass function is calculated by solving the one-body CST-Dyson equation, directly from the same kernel entering the two-body CST-Bethe-Salpeter equation. This makes our model completely self-consistent and in line with the traditional Dyson-Schwinger approaches. Other advances presented here are: (i) the proof of chiral symmetry breaking in a charge-conjugation symmetric CST 
four channel formulation, (ii) a new extended definition of the relativistic kernels that is needed when both particles are off-shell, essential for some applications. 

Before turning to the details of the model, we remind  the reader of an essential feature of the CST approach: quarks can have real mass poles and can be on their mass-shells.  Confinement is achieved by constructing an interaction that  guarantees that {\it two\/} or more quarks cannot be on-shell together.  The alternative view is that confinement occurs because the quark propagator does {\it not have any real mass-shell poles\/}.  For those who find the concept of an on-shell quark distasteful, it may help to think of the on-shell quark as a {\it effective\/} degree of freedom that is not physical, since, unlike nucleons (for example), quarks can never be isolated and can never be observed.  The issue is whether or not such an effective degree of freedom is useful in explaining the phenomena of QCD; until it is developed and tested a definitive answer cannot be given.   A feature of this  picture of confinement of quarks is 
that it is similar to the usual nonrelativistic picture, allowing us to make comparisons with nonrelativistic models. 

In Section II we motivate and write the charge-conjugation-invariant CST four channel equations. In Section III we focus on the
equations for the pseudoscalar bound state. In Section IV we prove that in the chiral limit the one-body CST-Dyson equation coincides with the two-body CST-Bethe-Salpeter equation corresponding to a zero mass Goldstone boson. Section V 
describes the general form of the CST kernel.  The results for the quark mass function and their connection to LQCD are shown in Section VI. Section VII summarizes and describes our conclusions.

A companion paper (prepared at the same time) uses the results of this paper to calculate the pion electromagnetic form factor  \cite{pion_form_factor_paper}.  It is referred to as Ref.~II.

\section{Charge conjugation invariant equations for quark-antiquark bound states}

\subsection{Background}

The purpose of this section is to motivate the structure of the equations used in this and forthcoming papers. The details of the interaction kernel are not of importance at this stage and will therefore be specified later in Section V.  

We begin  by clarifying and reviewing the relation between the Bethe-Salpeter (BS)~\cite{Sal51} and the CST equations.  The CST equations can be conveniently obtained from the BS equations by integrating over the internal energy variables and retaining only the contributions from certain propagator poles (for a two-body equation for non-identical particles with unequal mass, only the heavier particle pole is retained).   While this is a convenient technique for obtaining the equation, one might conclude from it that  the CST equation is merely the result of an  approximation of the BS equation.  Unfortunately, this interpretation misses the key point, namely that the omission of the poles of the kernel or other propagators reflects, in some cases, cancellations between various parts of the complete kernel and its iterations, in particular between ladder and crossed-ladder diagrams. One has to keep in mind that an exact BS kernel contains an infinite set of irreducible diagrams, which has to be truncated in 
practical calculations. When using a \emph{truncated} kernel (usually a ladder truncation), omitting its poles can actually yield a better approximation to the \emph{exact} BS equation (i.e., with a non-truncated complete kernel) than keeping them. This surprising observation plays a central role in the CST framework. Furthermore, the CST equations are manifestly covariant (in common with the BS equations) and, even when used with a truncated kernel (the ladder sum, for example), have a smooth nonrelativistic limit (not usually a feature of the BS equation with a truncated kernel). A brief review of the foundations of the CST and its many applications can be found in Ref.~\cite{Sta11}.

A unified description of all mesons composed of quark-antiquark pairs requires a bound-state equation that transforms correctly under charge conjugation: the equation for a bound-state particle should transform into the equation for the bound-state anti-particle.  In particular, both the vertex function and its charge conjugate must satisfy the same equation. For instance for the case of the pion, the vertex functions for \emph{both} $\pi^+$ and $\pi^-$, which are connected by charge conjugation, should be the solutions of the \emph{same} bound-state equation.

We begin the discussion of charge conjugation symmetry with the BS equation, which is manifestly covariant and naturally satisfies charge conjugation symmetry. As outlined above, the CST equation can be obtained from the BS equation by keeping only certain propagator poles of the BS integrand. As a consequence of the omission of some of the propagator poles, the CST equations  are not automatically charge conjugation symmetric. In most cases where this framework has been applied so far, in particular in the description of few-nucleon systems, charge conjugation invariance is not an important issue.  But for this paper, where we want to deal in particular with quark-antiquark systems of equal-mass quarks,  it is important.
In this section,  charge conjugation invariance is restored by symmetrizing the equations, leading to a system of four coupled equations~\cite{Savkli:1999me}. These are the ``four-channel CST equations.''

\begin{figure}[t]
\includegraphics[clip=0cm,width=9cm]{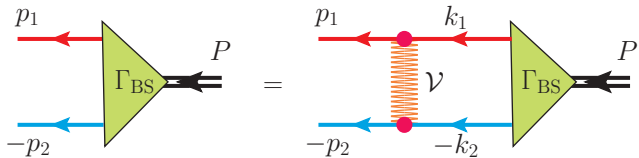}
\caption{ (Color online) The diagrammatic representation of the Bethe-Salpeter equation for the $q\bar q$ vertex function. The solid lines are the dressed propagators and the dots indicate the place where the Dirac space matrices ${\cal O}^i_j$ act. 
}\label{fig:1}
\end{figure}  

\subsection{Bethe-Salpeter equation}
\label{subsec:BSE}
The BS vertex function is denoted $\Gamma_\mathrm{BS}(p_1,p_2)$, with $p_1=p+\frac12P$ the four-momentum of the outgoing quark and $-p_2=-p+\frac12P$ the four-momentum of the outgoing antiquark (diagrammatically represented as an incoming quark of momentum $p_2=p-\frac12P$). $P$ is the bound-state four-momentum and $k_1=k+\frac12P$ and $-k_2=-k+\frac12P$  are the intermediate four-momenta of the quark and antiquark, respectively. With this notation the BS equation for the vertex function, shown diagrammatically in Fig.~\ref{fig:1}, is written 
\bea
\Gamma_\mathrm{BS}(p_1,p_2)&=&\mathrm i\int\frac{\mathrm {d}^4k}{(2\pi)^4}\,{\cal V}(p,k;P) 
\nonumber\\&&\qquad
\times S_1({k}_1)\,\Gamma_\mathrm{BS}(k_1,k_2)\,S_2(k_2)\,.\label{eq:BS}
\eea

Here ${\cal V}$ is the interaction kernel, which is an operator in the color and Dirac spaces of the two quarks whose interaction it describes. It is written in the general form
\begin{equation}
{\cal V}(p,k;P) = \frac {3}{4}{\bf F}_1 \cdot  {\bf F}_2 \sum_i V^i (p,k;P) {\cal O}^i_1 \otimes  {\cal O}^i_2 \, ,
\label{eq:BSkernelV}
\end{equation}
where ${\cal O}^i_1$ and ${\cal O}^i_2$ are Dirac matrices of type $i$ at the vertex involving quark 1 and 2, respectively, and the $V^i (p,k;P)$ are covariant scalar functions describing the corresponding momentum dependence. It is convenient to write the kernel in terms of the relative momenta $p, k$ and the total momentum $P$ instead of the individual particle labels, $p_1$, etc. How this kernel acts as an operator in Dirac space can be seen by following along a fermion line in Fig.\ \ref{fig:1}:
\begin{multline}
{\cal V}(p,k;P) \,S_1({k}_1)\,\Gamma_\mathrm{BS}(k_1,k_2)\,S_2(k_2) =  \\
 \sum_i V^i (p,k;P)  {\cal O}^i_1  \,S_1({k}_1)\, \Gamma_\mathrm{BS}(k_1,k_2)\,S_2(k_2) {\cal O}^i_2 \, .
\end{multline}

The color SU(3) generators are given in terms of the Gell-Mann matrices,  $F_a=\frac12\lambda_a$. 
Mesons are color singlet states, for which 
\begin{equation}
\langle {\bf F}_1 \cdot  {\bf F}_2 \rangle=\displaystyle\frac{4}{3} \,.
\end{equation}
The factor of $3/4$ in (\ref{eq:BSkernelV}) has been factored out in order to cancel this color matrix element. Color degrees of freedom can then be effectively ignored and will no longer be referred to in the remainder of this work. 

The dressed quark propagator, $S_i(k_i)$  (with the factor of $-\mathrm i$ removed), is given by
\bea
S_i(k_i)=\frac1{m_{0i}-\slashed{k}_i+\Sigma_i(\slashed{k}_i)-\mathrm i\epsilon} \label{eq:dressedprop}
\eea
with $m_{0i}$ the bare mass and $\Sigma_i$ the self-energy of the $i^{\rm th}$ quark, of the form
\bea
\Sigma_i(\slashed{k}_i)=A_i(k_i^2)+\slashed{k}_i B_i(k_i^2)\, .
\eea
If the quark and antiquark have the same bare mass and identical self interactions, the particle label on $S$ can be dropped,
\bea
S_1(k_i)=S_2(k_i)=S(k_i)\, .\label{eq:condprop}
\eea 

Charge conjugation, denoted by the operator ${\cal C}$, transforms quarks into antiquarks and vice versa, accomplished by taking the transpose of the vertex function and changing $p_1\leftrightarrow -p_2$.  The amplitude is invariant under charge conjugation if it remains unchanged up to a phase $\eta$, with $\eta^2=1$. The required condition is therefore
\bea
{\cal C}^{-1}\,\Gamma_\mathrm{BS}^T(p_1,p_2)\,{\cal C}=\eta \,\Gamma_\mathrm{BS}(-p_2,-p_1) \, , \label{eq:CC}
\eea
Performing this operation on Eq.~(\ref{eq:BS}), and using ${\cal C}^{-1}\,\gamma^{\mu T}\,{\cal C}=-\gamma^\mu$  and the charge conjugation invariant conditions 
\bea
{\cal C}^{-1}{\cal V}^T(p,k;P){\cal C}&=& {\cal V}(-p,-k;P)
\nonumber\\
{\cal C}^{-1}{S}^T(k){\cal C}&=& S(-k) \label{eq:symmcond}
\eea
where  $p_1=p+\frac12P\leftrightarrow -p_2=-p+\frac12P$  implies  $p\leftrightarrow-p$.  This gives
\begin{widetext}
 \bea
{\cal C}^{-1}\Gamma_\mathrm{BS}^T(-p_2,-p_1){\cal C}&=&\mathrm i\int\frac{\mathrm{d}^4k}{(2\pi)^4}\,{\cal V}(p,-k;P) \,S(-{k}_2)\,\Big[{\cal C}^{-1}\Gamma_\mathrm{BS}^T(k_1,k_2){\cal C}\Big]\,S(-k_1) 
\nonumber\\
&=&\mathrm i\int\frac{\mathrm {d}^4k}{(2\pi)^4}\,{\cal V}(p,k;P) \,S({k}_1)\,\Big[{\cal C}^{-1}\Gamma_\mathrm{BS}^T(-k_2,-k_1){\cal C}\Big]\,S(k_2)\, ,
\eea
\end{widetext}
which shows that ${\cal C}^{-1}\Gamma_\mathrm{BS}^T(-p_2,-p_1){\cal C}$ satisfies the same equation as $\Gamma_\mathrm{BS}(p_1,p_2)$  (and hence the two are equal up to a phase), provided conditions for the propagators and  kernel, Eqs.~(\ref{eq:condprop}) and (\ref{eq:symmcond}), are satisfied. 
We will always choose kernels that satisfy condition (\ref{eq:symmcond}). 

Note that a crucial step in the derivation was our ability to change the four-dimensional integration variable $k\to-k$.  This condition must be preserved when we specialize to the CST.

\begin{figure}[t]
\includegraphics[clip=0cm,width=8.5cm]{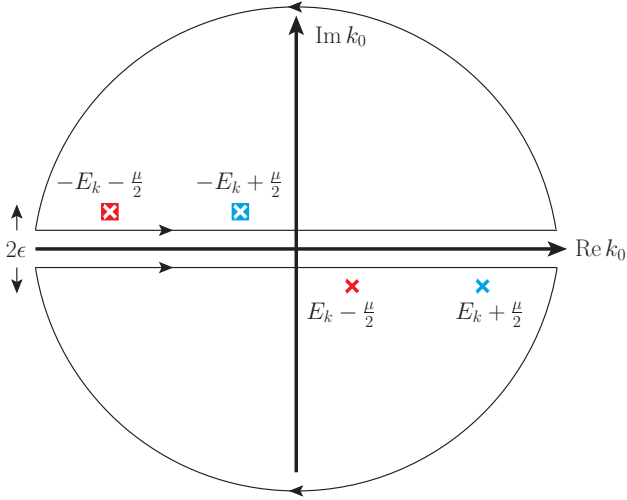}
\caption{(Color online) The positive-energy poles (colored crosses with positive $E_k$) and negative-energy (white crosses with negative $E_k$) poles of the propagators of quark 1 (red with $-\mu/2$) and quark 2 (cyan with $+\mu/2$) in the complex $k_0$-plane in the bound-state rest frame. }
\label{fig:poles1}
\end{figure} 

\subsection{Charge conjugation invariant CST equations}

Next we introduce a charge-conjugation invariant form of the bound-state CST equations. For cases when we want the correct limit as $P\to0$ these are the ``four-channel'' equations previously discussed~\cite{Savkli:1999me}.  

To motivate the structure of these equations, begin with the BS equation (\ref{eq:BS}) and consider the $k_0$ integration. 
The dressed propagator of quark $i$ with dressed mass $m$ and renormalization constant $Z_0$ can be written 
\bea
S(k_i)\simeq \frac{Z_0(m+\slashed{k}_i)}{m^2-k_i^2-\mathrm i\epsilon} \label{eq:Snearpole}
\eea
near its poles at $k_{i0}=\pm E_{k_i}$, where $E_{k_i} \equiv  \sqrt{m^2+{\bf k}_i^2}$.
Here we ignore the running of the dressed quark mass $m$ for the time being, but it will be included  later.
Figure \ref{fig:poles1} shows the positions of the four propagator poles in the complex $k_0$ plane in the bound-state rest frame (note that here $k_0$ is the zero component of the \emph{relative momentum} $k$, not of the individual particle momenta $k_i$). In the rest frame, the total momentum is $P_r=(\mu,{\bf 0})$, the quark and antiquark three-momenta ${\bf k}_{i}$ are equal to the relative three-momentum  ${\bf k}$, and therefore $E_{k_i}=E_k$, with $E_k \equiv \sqrt{m^2+{\bf k}^2}$. However, in the following we will continue working in an arbitrary frame with total momentum $P$ in order to emphasize the manifest covariance of our framework.

To perform the $k_0$ integration we can close the contour in the lower or upper half plane. In the CST framework only poles of propagators are included, whereas the poles of the kernel are moved to higher order kernels, which are usually neglected. As one can see in Fig.\  \ref{fig:poles1}, in either half plane the respective two poles are separated by the bound-state mass $\mu$. If $\mu$ is large, the pole closer to the origin dominates the integral, and the more distant pole can be neglected. However, in the limit $P \to 0$ the two poles move close together and the contributions of both must be taken into account.

First we close the $k_0$ contour in the lower half plane. Introducing the on-shell momenta $\hat k_{i}=(E_{k_i},{\bf k}_i)$ 
permits the two propagator pole contributions to the right hand side of (\ref{eq:BS}) to be written
 \bea
\Gamma(p_1,p_2)&=&-Z_0\int_{k_1}{\cal V}(p,\hat k_{1}-\sfrac12  P;  P)\Lambda(\hat k_1)
\nonumber\\&&\qquad\qquad\times
\Gamma( \hat k_1,\hat k_{1}- P)S(\hat k_{1}-  P)
\nonumber\\
&&
-Z_0\int_{k_2}{\cal V}(p,\hat k_{2}+\sfrac12  P;  P) S(\hat k_{2}+  P)
\nonumber\\&&\qquad\qquad\times
\Gamma(\hat k_{2}+  P,  \hat k_2)\Lambda(\hat k_2)\, ,\label{eq:two1a}
\eea
where $S(k_i)$ is the dressed propagator including the self-energy, and the positive-energy projection operator is
\bea
 \Lambda(\hat k)=\frac{(m+\slashed{\hat k})}{2m} \, ,
 \eea
and, for any $k$, the integral is abbreviated as
\bea
\int_k\equiv\int\frac{d^3k}{(2\pi)^ 3}\frac{m}{E_k}\, . \label{eq:kint}
\eea
Note that the notation used here for  $S(k_i)$ with the self-energy $\Sigma (k_i)$ is the same as in the previous section, but these are now CST objects and are not identical to their BS analogues.   For the remainder of the paper, they will always refer to CST quantities, unless otherwise stated. 

Equation~(\ref{eq:two1a}) can be simplified by renaming the integration variables ${\bf k}_i \rightarrow {\bf k}$ and using the shorthand notation 
$\hat k=(E_k,{\bf k})$ to denote the four-momentum of whichever particle is on-shell (so that, from this point on, ${\bf k}$ now denotes the three-momentum of the {\it on-shell\/} particle)

\begin{figure*}
    \includegraphics[width=1\textwidth]{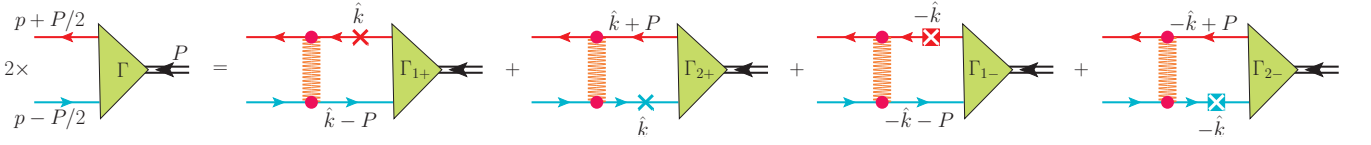}
\caption{ (Color online) The BS vertex function expressed in terms of the CST vertex functions. Note that here the antiquark of four-momentum $P-\hat p$ that propagates forward in time  (as in Fig.~\ref{fig:1}) has been interpreted as a quark of opposite four-momentum moving backward in time} \label{fig:BSVertexconst}
\end{figure*}

\begin{widetext}
\bea
\Gamma(p_1,p_2)=-Z_0\int_k\Big[&&{\cal V}(p,\hat k-\sfrac12  P;  P)\Lambda(\hat k)
\Gamma(\hat k,  \hat k- P)S(\hat k-  P)
+{\cal V}(p,\hat k+\sfrac12  P;  P) S(\hat k+  P)
\Gamma(\hat k+  P,  \hat k)\Lambda(\hat k)\Big]\, . \label{eq:two1}
\eea

\end{widetext}

We will now address the issue of charge conjugation symmetry of the CST bound-state equation. 
The equation just obtained is not charge conjugation invariant because it includes no poles from the upper half complex $k_0$-plane, necessary for the $k_0 \leftrightarrow -k_0$ symmetry used in the proof.  To correct this, we symmetrize the contributions from the contour in the lower half plane and the upper half plane, as illustrated in Fig.~\ref{fig:poles1}, which effectively amounts to taking their average. 

In the upper half plane, the sign for the energies $E_k$ is reversed. Guided by the argument used for the BS equation, we will also change the sign of the integration three-momentum $\bf k$, so that $\hat k\to -\hat k$.   The right hand side of the CST equation for the BS vertex now contains four terms (which will lead to the four-channel CST equation).  To simplify the notation, the kernel will be written ${\cal V}(p,k;P)\to {\cal V}(p,k)$, with the dependence on the total momentum $P$ understood, and we use the following abbreviations for the four amplitudes 

\begin{align}
&\Gamma_{1+}(k) \equiv  \Gamma(\hat k,\hat k-  P)  
\nonumber\\
& \Gamma_{2+}(k) \equiv \Gamma(\hat k+  P,  \hat k) 
\nonumber\\ 
&\Gamma_{1-}(k) \equiv  \Gamma(-\hat k,  -\hat k-P)  
\nonumber\\
& \Gamma_{2-}(k) \equiv \Gamma(-\hat k+  P,  -\hat k ) \, .
\label{eq:CSTamps}
\end{align} 

With this simplification, the equation becomes
\begin{widetext}
\bea
\Gamma(p_1,p_2)=-&&\frac12Z_0\int_k
\Big[{\cal V}(p,\hat k-\sfrac12  P)\Lambda(\hat k)\Gamma_{1+}(k)S(\hat k-  P)
+{\cal V}(p,\hat k+\sfrac12  P) S(\hat k+  P)\Gamma_{2+}(k)\Lambda(\hat k)
\nonumber\\&&
+{\cal V}(p,-\hat k-\sfrac12  P) \Lambda(-\hat k)\Gamma_{1-}(k)S(-\hat k-  P)
+{\cal V}(p,-\hat k+\sfrac12  P) S(-\hat k+  P)\Gamma_{2-}(k)\Lambda(-\hat k)\Big]\, ,
\nonumber\\  \label{eq:four1}
\eea

\end{widetext}
where we have replaced $\bold k$ by $-\bold k$ in the last two terms under the integral such that the negative-energy on-shell four-vector $(-E_k,\bold k)$ becomes equal to $-\hat k$. 
Note that the first and third, and the second and fourth terms differ only by $\hat k\to-\hat k$, preserving the symmetry required for charge conjugation invariance. 
 Equation (\ref{eq:four1}) is diagrammatically depicted in Fig.~\ref{fig:BSVertexconst}, and we will refer to it as the CST-BS equation. It expresses an approximate BS vertex function in terms of the CST vertex functions. Once the latter are known, it can be used to define a CST vertex function for states in which both quark and antiquark are off mass shell. But in order to determine the CST vertex functions in the first place, Eq.~(\ref{eq:four1}) needs to be converted into a closed set of four coupled equations, which is achieved by writing (\ref{eq:four1}) for each of the four values of the relative momentum $p\rightarrow\{\hat p-\frac12  P,\,\hat p+\frac12  P,\,-\hat p-\frac12  P,\,-\hat p+\frac12  P\}$ where $\hat p=(E_{p},{\bf p})$ is a individual quark on-shell momentum.  Introducing the 16 kernels
\begin{align}
&{\cal V}_{1\eta,1\eta'}(p,k)\equiv {\cal V}(\eta\hat p-\frac12 P,\eta'\hat k-\frac12P)
\nonumber\\
&{\cal V}_{1\eta,2\eta'}(p,k)\equiv {\cal V}(\eta\hat p-\frac12 P,\eta'\hat k+\frac12P)
\nonumber\\
&{\cal V}_{2\eta,1\eta'}(p,k)\equiv {\cal V}(\eta\hat p+\frac12 P,\eta'\hat k-\frac12P)
\nonumber\\
&{\cal V}_{2\eta,2\eta'}(p,k)\equiv {\cal V}(\eta\hat p+\frac12 P,\eta'\hat k+\frac12P)
\, . 
\end{align}
where a subscript $i\eta$ (with $\eta=\pm$) indicates that quark $i$ is on its positive or negative energy shell, respectively.
Note that a negative-energy quark should be interpreted as the corresponding physical positive-energy antiquark.  With this notation the four equations become
\begin{widetext}
\bea
\Gamma_{1+}(p)=-\frac12\int_k&&\Big[{\cal V}_{1+,1+}(p,k)\Lambda(\hat k)\Gamma_{1+}(k)S(\hat k-  P)
+{\cal V}_{1+,2+}(p,k) S(\hat k+  P)\Gamma_{2+}(k)\Lambda(\hat k)
\nonumber\\&&
+{\cal V}_{1+,1-}(p,k) \Lambda(-\hat k)\Gamma_{1-}(k) S(-\hat k-  P)
+{\cal V}_{1+,2-}(p,k) S(-\hat k+  P)\Gamma_{2-}(k)\Lambda(-\hat k)\Big]
\nonumber\\
\Gamma_{2+} (p)=-\frac12\int_k&&\Big[{\cal V}_{2+,1+}(p,k)\Lambda(\hat k)\Gamma_{1+}(k)S(\hat k-  P)
+{\cal V}_{2+,2+}(p,k) S(\hat k+  P)\Gamma_{2+} (k)\Lambda(\hat k)
\nonumber\\&&
+{\cal V}_{2+,1-}(p,k) \Lambda(-\hat k)\Gamma_{1-}(k) S(-\hat k-  P)
+{\cal V}_{2+,2-}(p,k) S(-\hat k+  P)\Gamma_{2-}(k)\Lambda(-\hat k)\Big]
\nonumber\\
\Gamma_{1-}(p)=-\frac12\int_k&&\Big[{\cal V}_{1-,1+}(p,k)\Lambda(\hat k)\Gamma_{1+}(k)S(\hat k-  P)
+{\cal V}_{1-,2+}(p,k) S(\hat k+  P)\Gamma_{2+} (k)\Lambda(\hat k)
\nonumber\\
&&+{\cal V}_{1-,1-}(p,k) \Lambda(-\hat k)\Gamma_{1-}(k) S(-\hat k-  P)
+{\cal V}_{1-,2-}(p,k) S(-\hat k+  P)\Gamma_{2+}(k)\Lambda(-\hat k)\Big]
\nonumber\\
\Gamma_{2-} (p)=-\frac12\int_k&&\Big[{\cal V}_{2-,1+}(p,k)\Lambda(\hat k)\Gamma_{1+}(k)S(\hat k-  P)
+{\cal V}_{2-,2+}(p,k) S(\hat k+  P)\Gamma_{2+} (k)\Lambda(\hat k)
\nonumber\\
&&+{\cal V}_{2-,1-}(p,k)\Lambda(-\hat k)\Gamma_{1-}(k) S(-\hat k-  P)
+{\cal V}_{2-,2-}(p,k)S(-\hat k+  P)\Gamma_{2-}(k)\Lambda(-\hat k)\Big] \, .\qquad
\label{eq:CST4ch}
%
\eea
\end{widetext}
The system of equations (\ref{eq:CST4ch}) is the four-channel CST equation.

The charge conjugation conditions (\ref{eq:CC}) are converted into connections between these amplitudes, namely
\bea
\Gamma_{1+}(p)&=&\eta\,{\cal C}^{-1}\Gamma^T_{2-}(p){\cal C};\quad  \Gamma_{2+}(p)=\eta\,{\cal C}^{-1}\Gamma^T_{1-}(p){\cal C}
\nonumber\\
\Gamma_{1-}(p)&=&\eta\,{\cal C}^{-1}\Gamma^T_{2+}(p){\cal C};\quad  \Gamma_{2-}(p)=\eta\,{\cal C}^{-1}\Gamma^T_{1+}(p){\cal C}\, . 
\nonumber \\ 
\label{eq:C1} 
\eea
Relations (\ref{eq:C1}) are consistent only if $\eta=\pm1$. 
The invariance of the four coupled-channel CST equations (\ref{eq:CST4ch}) under the substitutions (\ref{eq:C1}) is shown explicitly in the Appendix.

\section{Equations for the pseudoscalar bound state}

The pion as the lightest of the mesons requires a treatment consistent with chiral symmetry. In particular, we will show that, in the chiral limit of vanishing bare quark mass $m_0$, the pion bound state mass $\mu$ also tends to zero, while the constituent quark mass acquires a finite value due to its dressing through the interaction kernel. Our demonstration will be done in two steps:

In this section, we apply the four-channel CST equations (\ref{eq:CST4ch}) to the case of a pseudoscalar bound state and perform the  $\mu\to0$ limit. Then, in the next section, the resulting equation is shown to be consistent with the one-body CST-Dyson equation (i.e. the equation for the quark self-energy) in the chiral limit. This property is essential for the existence of a zero-mass pion solution, which plays the role of the Goldstone boson associated with the spontaneous breaking of chiral symmetry.

\subsection{General results}

The most general form for the BS vertex function for a pseudoscalar bound state can be written
\bea
&&\Gamma_{\rm BS}(p_1,p_2)= G_1(p_1^2,p_2^2)\gamma^5+G_+(p_1^2,p_2^2)(\slashed{p}_1\gamma^5+\gamma^5\slashed{p}_2)\nonumber\\&&\quad+G_-(p_1^2,p_2^2)(\slashed{p}_1\gamma^5-\gamma^5\slashed{p}_2)+G_3(p_1^2,p_2^2)\slashed{p}_1\gamma^5\slashed{p}_2,\quad
\label{eq:pion}
\eea
where $G_1$, $G_\pm$, and $G_3$ are scalar functions. 
If the state is invariant under charge conjugation with a phase $\eta$, then the relation (\ref{eq:CC}), together with $\left[\mathcal{C},\gamma^5 \right]=0$, $\mathcal{C}^{-1}  \gamma^{\mu T} \mathcal{C} = -\gamma^\mu$, and $\gamma^{5T}=\gamma^5$, leads immediately to the conditions 
\bea
G_1(p_1^2,p_2^2)&=&\eta G_1(p_2^2,p_1^2)\,;\nonumber\\ G_\pm(p_1^2,p_2^2)&=&\pm\eta G_\pm(p_2^2,p_1^2)\,;\nonumber\\ G_3(p_1^2,p_2^2)&=&\eta G_3(p_2^2,p_1^2)\, .
\eea
Using the decomposition (\ref{eq:pion}), the CST amplitudes (\ref{eq:CSTamps}) are
\begin{widetext}
\bea
\Gamma_{1+}(p)&=&G_1(m^2,p_-^2)\gamma^5 - G_+(m^2,p_-^2)\gamma^5\slashed P -G_-(m^2,p_-^2) \gamma^5(2\slashed{\hat p}-\slashed P)-G_3(m^2,p_-^2)(m^2\gamma^5+\slashed{\hat p}\gamma^5\slashed P)
\nonumber\\
\Gamma_{2+}(p)&=&G_1(p_{+}^2,m^2)\gamma^5 -G_+(p_{+}^2,m^2)\gamma^5\slashed P +G_-(p_{+}^2,m^2)(2\slashed{\hat p}+\slashed P)\gamma^5 -G_3(p_{+}^2,m^2)(m^2\gamma^5-\slashed P\gamma^5\slashed{\hat p})
\nonumber\\
\Gamma_{1-}(p)&=&G_1(m^2,p_+^2)\gamma^5 - G_+(m^2,p_+^2)\gamma^5\slashed P -G_-(m^2,p_+^2)(2\slashed{\hat p}+\slashed P)\gamma^5-G_3(m^2,p_+^2)(m^2\gamma^5-\slashed{\hat p}\gamma^5\slashed P)
\nonumber\\
\Gamma_{2-}(p)&=&G_1(p_-^2,m^2)\gamma^5 -G_+(p_{-}^2,m^2)\gamma^5\slashed P +G_-(p_{-}^2,m^2)\gamma^5(2\slashed{\hat p}-\slashed P) -G_3(p_{-}^2,m^2)(m^2\gamma^5+\slashed P\gamma^5\slashed{\hat p}) \, ,
\label{eq:CSTpionvertex1}
\eea
where  $p_{\pm}\equiv P\pm \hat p$. The vertex functions (\ref{eq:CSTpionvertex1}) satisfy the charge conjugation conditions (\ref{eq:C1}). Substituting the definitions (\ref{eq:CSTpionvertex1}) into (\ref{eq:four1}) and writing the result in the form (\ref{eq:pion}) gives four 
equations for the scalar BS vertex functions of the pion in terms of the CST vertex functions, which can be computed by substituting (\ref{eq:CSTpionvertex1}) into (\ref{eq:CST4ch}).  We will not pursue these general relations at this time; 
instead we look at the mass zero limit of the pion bound-state equation. 

\subsection{Zero-mass limit}\label{sec:masszerolimit}

For this case it is advantageous to go first to the bound-state rest frame where $P=P_r$, and then perform the limit $\mu\rightarrow0$. In the rest frame the vertex functions~(\ref{eq:CSTpionvertex1}) become
\bea
\Gamma_{1+}(p)&=&G_1(m^2,p_-^2)\gamma^5 -\mu G_+(m^2,p_-^2)\gamma^5\gamma^0 -G_-(m^2,p_-^2) \gamma^5(2\slashed{\hat p}-\mu\gamma^0)-G_3(m^2,p_-^2)(m^2\gamma^5+\mu\,\slashed{\hat p}\gamma^5\gamma^0)
\nonumber\\
\Gamma_{2+}(p)&=&G_1(p_+^2,m^2)\gamma^5 -\mu G_+(p_+^2,m^2)\gamma^5\gamma^0 +G_-(p_+^2,m^2)(2\slashed{\hat p}+\mu\gamma^0) \gamma^5-G_3(p_+^2,m^2)(m^2\gamma^5-\mu\,\gamma^0\gamma^5\slashed{\hat p})
\nonumber\\
\Gamma_{1-}(p)&=&G_1(m^2,p_+^2)\gamma^5 -\mu G_+(m^2,p_+^2)\gamma^5\gamma^0 -G_-(m^2,p_+^2)(2\slashed{\hat p}+\mu\gamma^0)\gamma^5-G_3(m^2,p_+^2)(m^2\gamma^5-\mu\,\slashed{\hat p}\gamma^5\gamma^0)
\nonumber\\
\Gamma_{2-}(p)&=&G_1(p_-^2,m^2)\gamma^5 -\mu G_+(p_-^2,m^2)\gamma^5\gamma^0 +G_-(p_-^2,m^2)\gamma^5(2\slashed{\hat p}-\mu\gamma^0)-G_3(p_-^2,m^2)(m^2\gamma^5+\mu\,\gamma^0\gamma^5\slashed{\hat p})\,. \label{eq:CSTpionvertex}
\eea
\end{widetext}

It will be shown shortly that the zero-mass limit of the BS vertex function for the pion can be obtained from the zero-mass  limit of the pion CST vertex functions (\ref{eq:CSTpionvertex}).  Since the pion is even under charge conjugation ($\eta=1$), $G_-=0$ when $\mu\to0$, and all of the vertex functions reduce to a single function, which will be denoted $\Gamma_0$, with 
\bea
\Gamma_{0}\equiv G_0\gamma^5=(G_{10}-m^2G_{30})\gamma^5
\eea
where $G_i(m^2,m^2)\equiv G_{i0}$.  Substituting this into (\ref{eq:four1}) for the BS vertex function gives
%
%
\bea
&& \Gamma(p,p)=[G_1(p^2,p^2)-p^2G_3(p^2,p^2)]\gamma^5\equiv G(p^2)\gamma^5
\nonumber\\
&&\quad=-\frac12G_0Z_0\int_k\Bigg\{{\cal V}(p,\hat k)\Theta(\hat k)
+{\cal V}(p,-\hat k) \Theta(-\hat k)\Bigg\}\, , \qquad\label{eq;BSpion0}
\eea
%
 where
\bea
\Theta(\hat k)\equiv \Lambda(\hat k)\gamma^5 S(\hat k-\mu)
+S(\hat k+\mu)\gamma^5 \Lambda(\hat k).\qquad
\eea
Here the $\mu\to0$ limit has been taken everywhere but in the propagators, where the numerators and denominators are both of order $\mu$, canceling to give a finite result in the $\mu\to0$ limit.  

To work out this limit and to evaluate $\Gamma(p,p)$, use the general form (\ref{eq:dressedprop}) of the dressed propagator, which can be written
\bea
S(k)=\left(\frac{1}{1-B}\right)\frac{M+\slashed{k}}{M^2-k^2-\mathrm i\epsilon}
= \frac{Z(M+\slashed{k})}{M^2-k^2-\mathrm i\epsilon} \qquad
\eea
where the mass function is
\bea
M=\frac{m_0+A}{1-B} \label{eq:massfunc}
\eea
and $Z$ is the wave function renormalization, and $A$, $B$, and $M$ are functions of $k^2$.  The dressed mass $m$ is defined by the condition 
\bea
M(m^2)=m=m_0+A_0+mB_0\, , \label{eq:massconstraint}
\eea
with $A_0=A(m^2)$ and $B_0=B(m^2)$.  Similarly, we define $Z_0=Z(m^2)$. With our application in mind, we expand the propagators about the energy of the four-vector $\hat k$, with $\hat k^2=m^2$ and
\bea
k_\pm=\hat k\pm P=\{E_k\pm\mu,\;{\bf k}\}\, .
\eea
Expanding $B$ and the mass function about $\hat k$ (to order $\mu$ is sufficient) gives
\bea
k^2_\pm&\simeq& m^2 \pm2\mu E_k\,,\quad\nonumber\\
M&\simeq& m\pm2\mu E_k M'\,, \quad\nonumber\\ B&\simeq& B_0\pm2\mu E_k B'\,,
\eea
where 
\bea
M'\equiv\frac{dM(k^2)}{dk^2}\Big|_{k^2=m^2}=\frac{A'+mB'}{(1-B_0)}
\eea
(with $A'$ and $B'$ defined similarly).  To order $\mu^0$ the dressed quark propagator becomes (with $\alpha=1$ for use with the first two terms in (\ref{eq;BSpion0}) and $-1$ for use with the last two) 
%
\bea
S(\alpha k_\pm)&\simeq&\frac{1}{1-B_0}\Big\{(m+\alpha\slashed{\hat k})\frac{1\pm4\mu E_kB'/(1-B_0)}{\pm2\mu E_k(2mM'-1)} 
\nonumber\\
&&+\frac{ 2\mu E_k M'+\alpha\slashed{P}}{2\mu E_k(2mM'-1)}+{\cal O}(\mu)\Big\} 
\nonumber\\
&\simeq&\frac{Z_0(m+\alpha\slashed{\hat k})}{m^2-k_\pm^2-i\epsilon}+(m+\alpha\slashed{\hat k})\frac{2Z_0B'}{(1-B_0)} 
\nonumber\\
&&+\frac{Z_0}{2E_k}(2E_kM'+\alpha\gamma^0)\, ,
\label{eq:Sexpansion}
\eea
where  $Z_0^{-1}=(1-B_0)(2mM'-1)$ is the renormalization constant.  The singular term and the $B'$ correction cancel when inserted into $\Theta$ [because $(m+\alpha\slashed{\hat k})\gamma^5(m+\alpha\slashed{\hat k})=0$], leaving finite terms 
\bea
\Theta(\alpha\hat k)&&\to
\Lambda(\alpha\hat k)\gamma^5S(\alpha k_-)+S(\alpha k_+)\gamma^5\Lambda(\alpha\hat k)
\nonumber\\
&&\to\frac{Z_0}{2E_k}\Big\{\Lambda(\alpha\hat k)\gamma^5 (2E_k M' + \alpha\gamma^0) 
\nonumber\\
&&\qquad+(2E_k M' + \alpha\gamma^0)\gamma^5\Lambda(\alpha\hat k)\Big\}
\nonumber\\
&&=\frac{Z_0}{2E_k}\gamma^5\Big(2E_k M'-\frac{E_k}{m}\Big)=\frac{\gamma^5}{2m(1-B_0)}\, . \quad\label{eq:cancellation}
\eea
Inserting this into (\ref{eq;BSpion0}) gives  finally
\bea
\Gamma(p,p)=-&&\frac{G_0 Z_0}{4m(1-B_0)}
\nonumber\\&&\qquad\times\int_k
\Big\{{\cal V}(p,\hat k)\gamma^5+{\cal V}(p,-\hat k) \gamma^5\Big\}\, .\qquad \label{eq;BSpion1}
\eea

Using the representation (\ref{eq:BSkernelV}) for a particular kernel with Lorentz scalar and vector couplings,
\bea
{\cal V}(p,k)&=& V_S(p,k){\bf 1}_1\otimes{\bf 1}_2+\frac14V_V(p,k)\mathrm g_{\mu\nu}\bm{\gamma}_{1}^\mu\otimes{\bm \gamma}^{\nu}_2 \qquad\label{eq:BSkernel}
\eea
shows that
\bea
G(p^2)=\frac{G_0Z_0}{4m(1-B_0)}\int_k
\bigg[&& V_V(p,\hat k)+{V}_V(p,-\hat k)
\nonumber\\
&&-{V}_S(p,\hat k)-{V}_S(p,-\hat k)\bigg].\qquad
\label{eq;BSpion2}
\eea
Since $G(m^2)=G_0$, this equation has a non-zero solution only if
\bea
1=\frac{Z_0}{4m(1-B_0)}\int_k
\bigg[&&V_V(\hat p,\hat k)+{V}_V(\hat p,-\hat k)
\nonumber\\
&&-{V}_S(\hat p,\hat k)-{V}_S(\hat p,-\hat k)\bigg].\qquad
\label{eq:pionconstraint}
\eea
This equation is the condition for the existence of a massless solution of the pion bound-state equation and will be referred to as the massless pion condition. 

\section{Chiral-symmetry breaking in the CST}
In the GMS model the spontaneous breaking of chiral symmetry is realized through the famous NJL-mechanism: In the chiral limit of vanishing bare quark mass chiral symmetry is dynamically broken by a finite dressed quark mass that is spontaneously generated by the self-interactions of the quark through the interaction kernel. This symmetry breaking is accompanied by the existence of a zero-mass pseudoscalar bound-state, the Goldstone pion: It can be shown analytically that the one-body CST-Dyson equation for the scalar self-energy becomes identical to the two-body CST-BS equation for a massless pion as $m_0\rightarrow0$. Therefore, the existence of the solution for one implies a solution for the other. On the other hand, away from the chiral limit, the existence of a solution of the one-body equation ensures that there cannot exist a massless-pion solution of the two-body equation at the same time.

In this section we derive the condition for the scalar part of the CST one-body equation to have a non-trivial solution and therefore to generate a contribution to the dressed quark mass. In the chiral limit this condition reduces to~(\ref{eq:pionconstraint}) which guarantees the existence of a massless pion, and at the same time ensures that the pion is not massless for finite bare quark masses.   It will be seen that for this mechanism to work the interaction kernel must satisfy certain conditions. In particular, for the case of the kernel of Eq.~(\ref{eq:BSkernel}), $V_S$ must \emph{not} contribute to either the one-body equation for the scalar self-energy or to the two-body equation for a massless pion.

\subsection{Mass functions}\label{sec:massfun}

\begin{figure}
\includegraphics[clip=0cm,width=8.5cm]{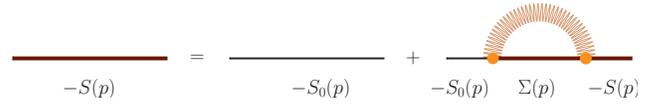}
\caption{(Color online) The diagrammatic representation of the one-body Dyson equation for the self-energy. The thick line denotes the dressed propagator, the thin line the bare propagator.
}\label{fig:DE}
\end{figure}

Using the kernel~(\ref{eq:BSkernelV}), the BS self-energy 
\bea
\Sigma(p)&=&-\mathrm i\int\frac{\mathrm d^4k}{(2\pi)^4} {\cal V}(p,k) S(k)\nonumber\\&=&  -i\int\frac{d^4k}{(2\pi)^4}\frac{{\cal V}(p,k) \big[M+\slashed{k}\big]}{(1-B)(M^2-k^2-i\epsilon)}\,
\eea
is given from the four-dimensional one-body Dyson equation diagrammatically depicted in Fig.~\ref{fig:DE}.   (Here, since  $P=0$, the  
$p$ and $k$ are  \emph{individual} quark off-shell four-momenta.)

\begin{figure*}
\includegraphics[clip=0cm,width=17
cm]{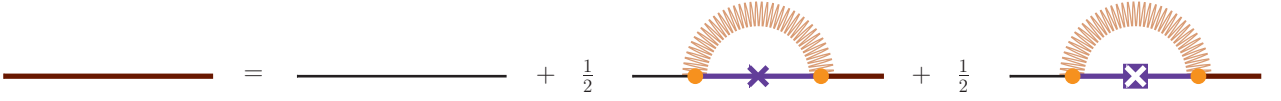}
\caption{(Color online) The diagrammatic representation of the one-body CST-Dyson equation for the self-energy. 
}\label{fig:DECST}
\end{figure*}

In the CST, we are instructed to take an average of the contributions from the quark pole in the lower half plane, where the singularity is at $k_0=E_k$, and the upper-half plane where the singularity is at $k_0=-E_k$, see Fig.~\ref{fig:poles1}. The corresponding one-body CST-Dyson equation is diagrammatically depicted in Fig.~\ref{fig:DECST}. 


\begin{figure*}
\includegraphics[clip=0cm,width=10cm]{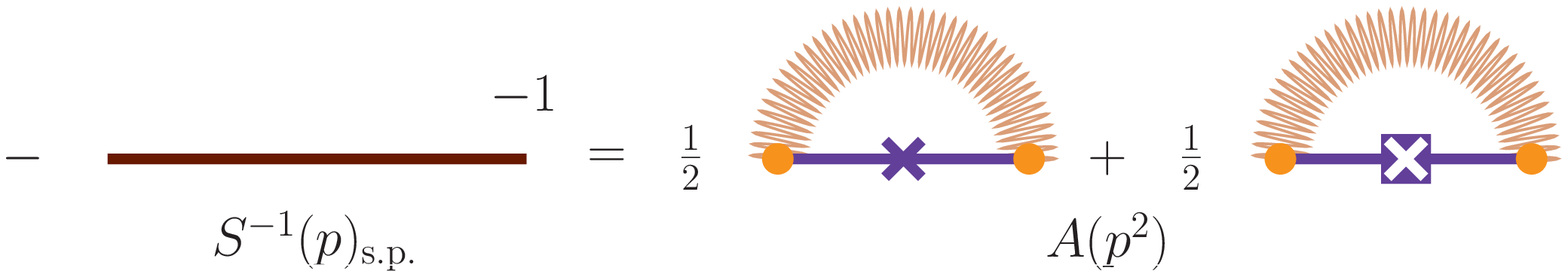}\\\vspace{0.5cm}
\includegraphics[clip=0cm,width=18cm]{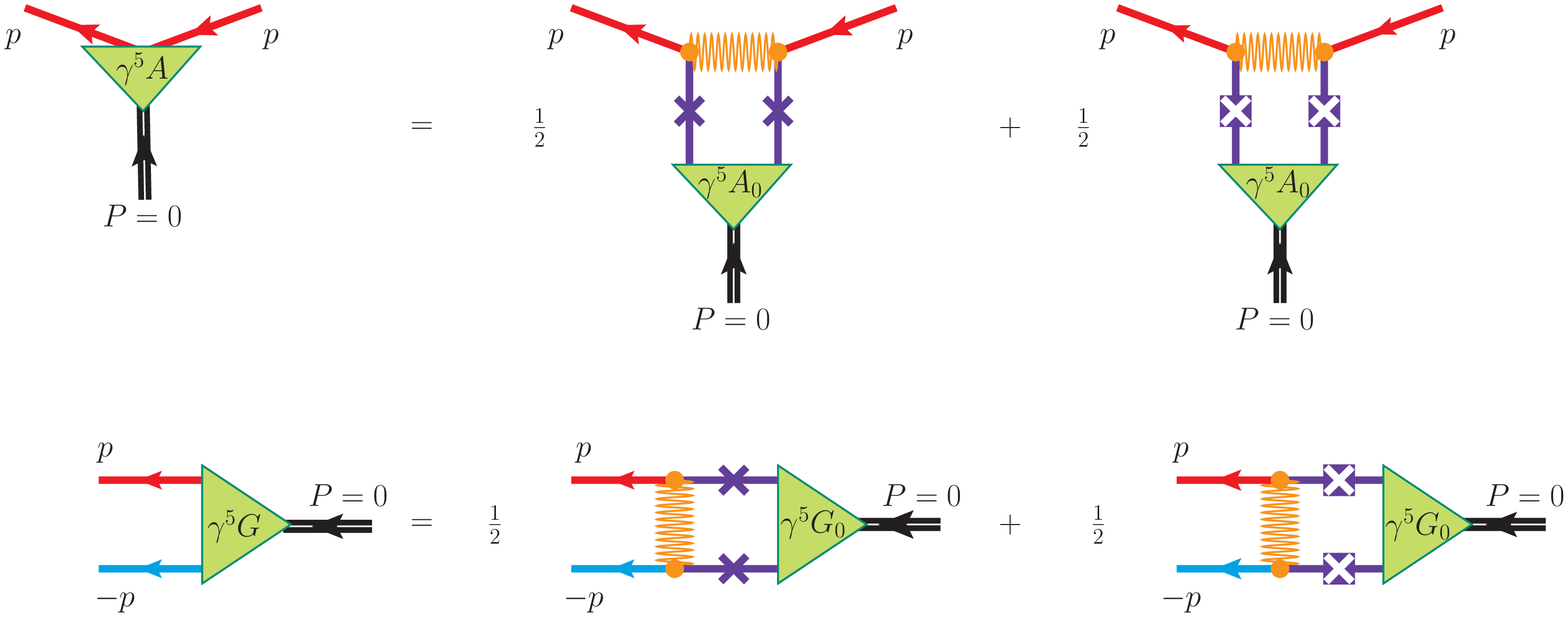}
\caption{(Color online) The equivalence between the BS bound-state vertex function for a zero-mass pion and the scalar part (s.p.) of the CST self-energy in the chiral limit.
}\label{fig:spCSTDE}
\end{figure*}

Calculating these pole contributions from the singular (first) term in the expansion (\ref{eq:Sexpansion}), and changing ${\bf k}\to-{\bf k}$ in the negative energy term (so that $\hat k\to-\hat k$) gives
\bea
\Sigma(p)=\frac12Z_0\int_k&&
\Big\{{\cal V}(p,\hat k)\Lambda(\hat k)+{\cal V}(p,-\hat k)\Lambda(-\hat k)\Big\}\,.\qquad
\label{eq:fullSigma}
\eea
Inserting  the kernel~(\ref{eq:BSkernel}) we obtain
\bea
\Sigma(p)=\frac{Z_0}{4m}\int_k&&
\Big\{m\big[V_S(p,\hat k)+V_S(p,-\hat k)
\nonumber\\&&
+V_V(p,\hat k)+V_V(p,-\hat k)\big]
\nonumber\\
&&+\slashed{k}\big[V_S(p,\hat k)-V_S(p,-\hat k)
\nonumber\\&&
-\frac12V_V(p,\hat k)+\frac12V_V(p,-\hat k)\big]\Big\}\, .\qquad\label{eq:CSTgapE}
\eea
For simplicity, evaluate this in the rest frame (${\bf p}=0$), where the integral over $\mathrm d^3k$ ensures that $\slashed{\hat k}\to \gamma^0 E_k$, and extract the self-energy functions $A$ and $B$
\bea
A(p_0^2)=\frac{Z_0}{4} \int_k&& \big[V_S(p,\hat k)+V_S(p,-\hat k)
\nonumber\\&&
+V_V(p,\hat k)
+V_V(p,-\hat k)\big]
\nonumber\\
B(p_0^2)=\frac{Z_0}{4p_0}\int_k&& \left[\frac{E_k}{m}\right] \big[V_S(p,\hat k)-V_S(p,-\hat k)
\nonumber\\&&
-\frac12V_V(p,\hat k)+\frac12V_V(p,-\hat k)\big]\, .
\label{eq:AandB}
\eea
Note that, since ${\bf p}=0$, $A$ and $B$ are functions of $p^2=p_0^2$, as required by Lorentz covariance.

At $p=\hat p$ (so that $p_0=m$), the constraint (\ref{eq:massconstraint}) gives
\bea
m(1-B_0)=m_0+A_0\, .
\eea
Using this reduces the equation for $A$ at $p_0=m$ to the constraint
\bea
1=\frac{1+m_0/A_0}{4m(1-B_0)}Z_0 \int_k \big[&&V_S(\hat p,\hat k)+V_S(\hat p,-\hat k)
\nonumber\\&&
+V_V(\hat p,\hat k)+V_V(\hat p,-\hat k)\big]\, . \qquad\label{eq:Aconstraint}
\eea
Comparing Eqs.~(\ref{eq:pionconstraint}) and (\ref{eq:Aconstraint}) shows that they are identical if $m_0=0$ and if the integral over the scalar interaction vanishes:
\bea
\int_k \big[V_S(p,\hat k)+V_S(p,-\hat k)\big]=0\,.\label{eq:condV_L}
\eea
This condition is satisfied by the models discussed in this paper.

Unless $m_{0}=0$, constraint (\ref{eq:pionconstraint}) will not be satisfied, insuring that there is no pion bound state with zero mass. This means that the same constraint that makes it possible that $m\ne0$ [Eq.~(\ref{eq:Aconstraint})] also ensures that there exists a pion with zero mass [Eq.~(\ref{eq:pionconstraint})] in the chiral limit. These consistency conditions link the spontaneous generation of a dressed quark mass in the chiral limit, and hence the spontaneous breaking of chiral symmetry, to the existence of a massless Goldstone pion. 

The equivalence between the zero-mass pion equation and the self-energy equation in the chiral limit can also easily be demonstrated in terms of Feynman diagrams, as shown in  Fig.~\ref{fig:spCSTDE}: The scalar self-energy $A(p^2)$ becomes equal to the scalar part of the inverse dressed propagator in the chiral limit. Multiplying with a $\gamma^5$, attaching two off-shell quark lines of momentum $p$ and one pion line of zero momentum yields a \lq \lq spacelike'' Yukawa vertex. Then, crossing symmetry allows to switch to the \lq\lq timelike'' creation/annihilation channel, which is equivalent to the (approximated) BS pion vertex function (expressed in terms of CST functions) in the chiral limit. This shows that, in the chiral limit, the scalar part of the quark self-energy, $A(p^2)$, is equivalent to $G(p^2)$, the BS vertex function for an off-shell outgoing quark and antiquark with momenta $p$ and $-p$, respectively.

A word of caution about the interpretation of Fig.~\ref{fig:spCSTDE} is in order:  while the drawing shows that the two internal quarks are both on-shell at the same time, this contribution does {\it not\/} correspond to quark-antiquark scattering.  As already  shown in (\ref{eq:cancellation}), the singular part of the dressed propagator, which is the only part that would contribute  to true scattering and deconfinement \cite{Savkli:1999me}, cancels.   Hence  the contribution to the diagram is the finite, non-singular part, not associated with any propagation of the quarks.

\section{General definition of the kernel}
Our discussion so far on charge conjugation and chiral-symmetry breaking applies to all covariant interaction kernels of the form (\ref{eq:BSkernel}) that satisfy the symmetry property (\ref{eq:symmcond}) together with the condition on the scalar part $V_S$, Eq.~(\ref{eq:condV_L}). In this section we specify the momentum-depended parts of the kernel. They include confinement and they are manifestly covariant expressions with the correct nonrelativistic limit, a property crucial for a unified description of all mesons.
\subsection{The relativistic kernel}\label{subsec:LSkernel}

The kernel we use is a relativistic generalization of the nonrelativistic potential
\bea
V^{\mathrm{nr}}(r)=&& V_L^\mathrm{nr}(r)+V_C^\mathrm{nr}
\nonumber\\
=&&\sigma r -C \label{eq:Vnr}
\eea
which is the superposition of a linear confining interaction, $V_L^\mathrm{nr}$, and a constant interaction, $V_C^\mathrm{nr}$.  
For a particular choice of the parameters $\sigma$ and $C$, this potential is similar to the very successful Cornell potential~\cite{Eichten:1975,Eichten:1978,Richardson:1978bt} used for the description of the charmonium spectrum. The constant interaction is used for simplicity. It can be replaced by a color-Coulomb potential if needed.

\begin{figure}
\includegraphics[clip=0cm,width=8.5cm]{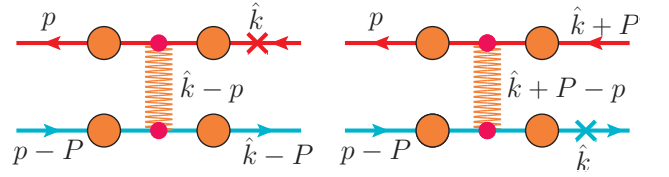}
\caption{(Color online) Two possible relativistic generalizations of the interaction kernel. For the particular case when $p$ is a on-shell momentum, i.e. $p=\hat p$, then the left panel describes the diagonal and the right panel describes the off-diagonal part of the kernel.  The orange blobs on the quark lines denote strong quark form factors.
}\label{Fig1b}
\end{figure}

A proper CST generalization of~(\ref{eq:Vnr}) to be applicable to the four-channel equations~(\ref{eq:four1}) must include the condition that {\it either\/} the quark {\it or\/} the antiquark are on-shell in the initial state.  The kinematics for these two terms is illustrated in Fig.~\ref{Fig1b}.
The covariant kernels in momentum space will be denoted as $V_L$ and $V_C$.   In the rest frame of the bound state, where $\bold k_1=\bold  k_2=\bold k$, the operation of the confining part $V_L$ on any regular function $\phi (k)$ will be defined by the  result
\bea
\langle V_L\phi\rangle(p)=\int_k V_{A}(p,\hat k;P) \left[\phi(\hat k)-\phi(\hat p_R)\right] \label{eq:VLrelativistic}
\eea
where, if the momenta are labeled as in Fig.~\ref{Fig1b}, with $p_1=p$ and $p_2=p-P$, then the kernel in the left panel is 
\bea 
V_{A}(p,\hat k;P)= -h(p_1^2)&&h(p_2^2)h(\hat k_1^2)h(k_2^2)
\frac {8\pi\sigma}{(p-\hat k)^4} \, .
\label{eq:VA}
\eea
The $h$'s are dimensionless strong quark form factors to be discussed in Section~\ref{sec:strongff}.
The two cases $V_{A}(\hat p,\hat k)$ and  $V_{A}(\hat p,\hat k+P)$ are called \lq\lq diagonal'' and  \lq \lq off-diagonal'' parts of the kernel,  corresponding for on-shell $p=\hat p$ to the left and right panel in Fig.~\ref{Fig1b}, respectively. The fact that (\ref{eq:VLrelativistic}) does indeed describe a linear confining interaction in the non-relativistic limit has been extensively discussed \cite{Gross:1991te,Savkli:1999me}.  

The ``constant'' part of the kernel, $V_C$, is defined by 
\bea
\langle V_C\phi\rangle(p)&=&\int_k V_C(p,\hat k;P)\phi(\hat k)
\nonumber\\
&=&2C\,h(p_1^2)h(p_2^2)h(\hat k_1^2)h(k_2^2)\,\phi(\hat p)
\label{eq:VCrelativistic}
\eea
where 
\bea
V_C(p,\hat k;P)&=& 2C\, h(p_1^2)h(p_2^2)h(\hat k_1^2)h(k_2^2)
\nonumber\\&&\times (2\pi)^3
\frac{E_k}{m}\delta^3 (p-k)
\label{eq:VC}
\eea
and $m$ is the dressed quark mass.  Note that the sign in (\ref{eq:VC}) is consistent with the sign in the nonrelativistic limit (\ref{eq:Vnr}) because a vector interaction changes sign when taking the nonrelativistic limit \cite{Uzz99}.

The factor of $m/E_k$  contained in the volume integrals used in~(\ref{eq:VLrelativistic}) and~(\ref{eq:VCrelativistic}) 
ensures that the three-dimensional volume integrals are covariant (because ${\bf k}$ is the three-momentum of an on-shell quark of mass $m$), so that the whole expression is covariant if the $\phi$'s are covariant.   They also ensure a smooth transition to the nonrelativistic limit when $m\to\infty$. The subtraction term, $\phi(\hat p_R)$, is an improved version of the one proposed in Ref.~\cite{Gross:1991te}, with the subtraction generalized to properly regulate both the ``diagonal'' singularities of $V_{A}$  at $(\hat k-\hat p)^2=0$ and  the ``off-diagonal'' singularities  at $(\hat k+P- \hat p)^2=0$.   The argument of the subtraction term is $\hat p_R=(E_{p_R},\bold p_R)$, with $\bold p_R=\bold p_R(p_0,\bold p)$ the value for $\bold k$ at which $V_{A}(p,\hat k)$ becomes singular. 
For instance, for $\bold p=0$ and $P=0$ (which occurs in the evaluation of the quark self-energy in the quark rest frame -- see Sec.~\ref{sec:Mfucntion}), we have
\bea
\bold p_R^2=\frac{1}{4p_0^2}(m^2-p_0^2)^2\, .
\eea 
The subtraction term has two important functions: (i) it regularizes the singular confining kernel $V_A$,  reducing it to a unique and well-defined Cauchy principal value integral,
and (ii) it ensures that the relativistic generalization of the nonrelativistic condition $V_L^{\mathrm{nr}}(r=0)=0$, written in momentum space 
\bea
\langle V_L\rangle= \int_k V_L(p,\hat k)=0\,  .  \label{eq:cond}
\eea
is satisfied.

Although the covariant interaction kernel has been constructed in such a way that it reduces to the linear confining potential in the nonrelativistic limit, it is \textsl{a priori} not obvious that it actually confines. As already mentioned in the Introduction, there are, in principle, two ways confinement can be realized: (1)   the dressed propagators have no real quark mass poles, or (2) the bound-state vertex function vanishes as two or more quarks are simultaneously on-mass-shell. It has been proven in Ref.~\cite{Savkli:1999me} that the latter realization of confinement applies to the CST approach. For this to work out the subtraction term in Eq.~(\ref{eq:VLrelativistic}) is crucial as it becomes singular when both quarks are on-shell, which forces the vertex function to vanish, for consistency. 

This concludes our discussion of the momentum-dependent parts of the kernel and we  turn now to its Lorentz structure.

 \begin{figure*}\begin{center}  
\includegraphics[clip=0cm,width=13cm]{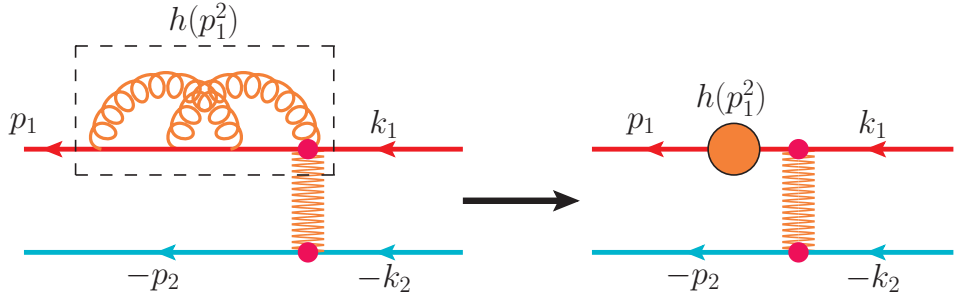}\end{center}
\caption{(Color online) The strong quark form factors can be viewed  as  gluon loop corrections to the quark vertices (shown here at the two-loop level).
}\label{fig:hff_vertices}
\end{figure*}

\subsection{Lorentz structure of the kernel}\label{eq:decoupl}
An appealing feature of the GMS model is the property that the linear confining interaction $V_L$ does not contribute to the CST equation for a massless pseudoscalar bound state~\cite{Gross:1991pk}. This can immediately be seen from Eq.~(\ref{eq;BSpion2}) by inserting $V_L$ for $V_S$ and $V_V$ in the kernel and using condition~(\ref{eq:cond}). The same applies to the CST equation for the scalar self-energy $A(p_0^2)$, Eq.~(\ref{eq:AandB}). Therefore, 
\bea 
A_L(p_0^2)=0 \label{eq:A_L=0}\,,
\eea
and hence the linear confining interaction does not contribute to the generation of a dressed quark mass. This decoupling of confinement from chiral symmetry breaking permits our confining potential to have a Lorentz structure that includes, e.g., a Lorentz-scalar coupling part as suggested from phenomenological approaches~\cite{Kalashnikova:2005tr,Bicudo:2003ji,Allen:2000sd} and LQCD calculations~\cite{Michael:1985rh,Gromes:1984ma,Bali:1997am,Lucha:1991vn}. Specifically, we employ a mixture of scalar and vector coupling, where $\lambda$ is the mixing parameter:
\bea
{\cal V}_L(p,\hat k)=\left[\lambda {\bf 1}_1\otimes{\bf 1}_2-(1-\lambda)\bm{\gamma}^\mu_1\otimes{\bm \gamma_\mu}_2 \right]V_L(p,\hat k) \label{eq:linearkernel}\,.\nonumber\\
\eea
The relative minus sign between scalar and vector parts guarantees that the kernel evaluated between on-mass-shell spinors for the quarks and antiquarks yields the correct potential~(\ref{eq:Vnr}) in the nonrelativistic limit, without renormalizing its strength $\sigma$.

Unlike the confining interaction, the constant interaction \emph{does} contribute to the equation for a zero-mass pseudoscalar bound-state. This constrains its Lorentz structure to be chirally symmetric. There are several possible choices, however not all of them also yield a zero-mass pion in the chiral limit~\cite{Gross:1991te}. In particular, if one allows only Lorentz scalar and vector structures as in (\ref{eq:BSkernel}), then condition (\ref{eq:condV_L}),  which must be satisfied in order to have  dynamical chiral-symmetry breaking, requires the scalar part to vanish.  Therefore we assume a pure vector structure for the constant interaction, which, in arbitrary color gauge and using the conventions of Eq.~(\ref{eq:BSkernel}), is written 
\bea
\mathcal V_C (p,\hat k)&=&\gamma^\mu_1 \otimes\gamma^\nu_{2} V_C (p,\hat k)
\nonumber\\&&\times\frac14\left[\mathrm g_{\mu\nu}-(1-
\xi)\frac{(p-k)_\mu(p-k)_\nu}{(p-k)^2}\right] \label{eq:V_C}\, ,\qquad
\eea
where $V_C$ was defined in Eq.~(\ref{eq:VC}).
Here $\xi$ is the gauge parameter. The choice $\xi=1$ corresponds to the Feynman gauge and $\xi=0$ to the Landau gauge. The constant $C$ included in $V_C$ will be chosen to be consistent with chiral symmetry and to give a reasonable description of the quark structure. 
\subsection{Strong quark form factors}\label{sec:strongff}

The kernels given in  Eqs.~(\ref{eq:VLrelativistic}) and~(\ref{eq:VCrelativistic})  include (strong) phenomenological form factors~\cite{Gro87,Gro92,Gross:1994he,Gro96}.   
Their purpose is twofold: they can be regarded as describing some gluonic corrections to the vertices that would otherwise be overlooked, and they provide convergence in loop integrals. The form factors are expressed as products of  individual quark form factors $h(p^2)$, one for each quark line associated with momentum $p$.

The idea that these form factors serve as an effective  description of the infinite sum of overlapping gluon loop corrections to the interaction vertices is illustrated in Fig.~\ref{fig:hff_vertices} (shown only for  two-loops).  This raises the possibility that they could be calculated from first principles at a later date.

The explicit form of $h(p^2)$ will be specified later. Our quark form factors differ from those used in previous approaches by their normalization at the on-shell point $p^2=m^2$.   They will be normalized to 1 only in the chiral limit, $h(m_\chi^2)=1$, and therefore $h(m^2)\neq1$. With this normalization all free parameters of the quark mass function are fixed in the chiral limit and the constituent quark mass $m$ for finite $m_0$ is then  uniquely determined from the mass constraint~(\ref{eq:massconstraint}).  

\begin{figure}\begin{center}  
\includegraphics[clip=0cm,width=8.5cm]{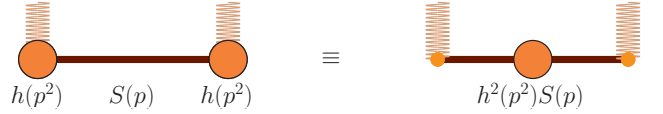}\end{center}
\caption{ (Color online) The strong quark form factors at the interaction vertices (left) or equivalently multiplied with the propagator (right).
}\label{fig:hff}
\end{figure}

The use of \emph{factorized} form factors [where the form factor at a vertex, $H(p,p')$, is separated into a product of separate form factors, $h(p^2)h(p'^2)$]  is an advantage when calculating electromagnetic current matrix elements in the presence of strong form factors.  If they are moved from the vertices to the  quark propagators (leaving the vertices bare) as illustrated in Fig.~\ref{fig:hff}, and reinterpreted as an additional correction to the quark propagator, then a dressed quark current can be constructed that includes them and also satisfies the Ward-Takahashi identity \cite{Gro87}. This is discussed in Ref.~II.

\section{Quark mass function}\label{sec:Mfucntion}

In this section we apply the interaction kernel defined in the previous section to calculate the CST self-energy given in Eq.~(\ref{eq:AandB}), and consequently the dynamical quark mass function $M(p^2)$ from Eq.~(\ref{eq:massfunc}). Our Minkowski-space results are then compared with LQCD calculations performed in Euclidean space. This comparison requires the computation of our mass function at negative values of $p^2$.  

\subsection{Self-energy from the linear confining potential}

The discussion on the decoupling of confinement from chiral symmetry breaking revealed that the linear confining potential does not contribute to the dynamical quark mass generation because the corresponding contribution to the scalar self-energy vanishes, as expressed in Eq.~(\ref{eq:A_L=0}). For the remaining vector part of the self-energy we insert the confining kernel (\ref{eq:VLrelativistic}) into Eq.~(\ref{eq:AandB}) and get
\bea
B_L(p_0^2)&=&\frac{(\lambda-2)}{2p_0} h^2(m^2)h^2(p_0^2)\nonumber\\&&\times \int\frac{\mathrm d^3k\,Z_0}{(2\pi)^3} \big[V_L(p,\hat k;P)-V_L(p,-\hat k;P)\big]\nonumber\\&=&-(\lambda-2)2\pi \sigma
\frac{ (p_0^2+m^2)}{p_0^4}h^2(m^2)h^2(p_0^2)\nonumber\\&&\times \int\frac{\mathrm d^3k\,}{(2\pi)^3}\frac{Z_0}{E_k^2-E_{p_R}^2} \, ,
\label{eq:B_L}
\eea
where the form factors $h(p^2)$ discussed in Sec.\ref{sec:strongff} have been introduced at each vertex. Equation (\ref{eq:B_L}) shows  that $B_L$ is a function of $p_0^2$, as it should be. Note that this would not be the case if we had not included the negative-energy propagator-pole contribution from the upper half complex $k_0$-plane.
Furthermore, (\ref{eq:B_L}) displays the simple dependence of $B_L$ on the adjustable mixing parameter $\lambda$.  For the particular choice $\lambda=2$, $B_L$ is zero and the confining potential does not contribute at all to the quark self-energy, i.e.,
\bea
\Sigma_L(p_0)=0\qquad \text{for}\qquad \lambda=2\,.
\eea
This choice corresponds to a $2:1$ ratio between scalar and vector coupling in the Lorentz structure of the confining kernel. 

The case $\lambda=2$ is appealing because of its simplicity as it avoids the computation of the UV-divergent integral in (\ref{eq:B_L}), and therefore we will explore this case in the remainder of this work. However, in future work we will compute the integral (\ref{eq:B_L}) using Pauli-Villars regularization, and $\lambda$ will be determined through a fit to the meson spectrum. It appears unlikely that $\lambda=2$ should emerge automatically from such a fit, and we expect that the confining interaction will then actually contribute to the quark self-energy.

\subsection{Self-energy and mass function from the constant interaction}
Next we look at the self-energy contribution $\Sigma_C$ from the constant potential $\mathcal V_C$ of Eq.~(\ref{eq:V_C}). Returning to Eq.~(\ref{eq:fullSigma}), in the rest frame of the quark, where its four-momentum is $(p_0,\bold 0)$, we find
\begin{eqnarray} \label{eq:se1}
 \Sigma_C(p_0)&=&
 \frac12 Z_0 \int_k 
\Big[\mathcal V_C(p_0,\hat k) \Lambda({\hat k})
\nonumber\\
&&\qquad
\qquad+\mathcal V_C(p_0,-\hat k) \Lambda(-{\hat k})\Big] 
\nonumber\\
&=&\left(\frac34+\frac14\xi\right)C\, 
h^2(p_0^2)
\,  h^2(m^2)\equiv A_C(p_0^2) \, ,\qquad
\label{eq:AC}
\end{eqnarray}
where ${\cal V}_C$ has been evaluated using Eqs.~(\ref{eq:VCrelativistic}) and (\ref{eq:V_C}), $Z_0=1$ (see below),  and $\xi$ is the  gauge parameter taken from (\ref{eq:V_C}).  Equation (\ref{eq:AC}) also shows that
\begin{eqnarray}
 B_C(p_0^2)=0\, ,
\end{eqnarray}
because the two pole contributions cancel. This result implies that a wave-function renormalization factor $Z(p^2)\neq1$ can only originate from the confining interaction. However,  since we chose $\lambda=2$, the confining interaction does not contribute to the self-energy at all and $Z(p^2)=1$, as assumed above. This fact has to be kept in mind when our mass function for $\lambda=2$ is compared to LQCD calculations, such as the ones of Ref.\ \cite{Bowman:2005vx}, where the mass function contains an overall factor $Z(p^2)$ substantially smaller than 1 at low $p^2$.

The mass function is not a physical observable and depends on the choice of the color gauge. In our case, this is explicit in Eq.~(\ref{eq:se1}) through the term containing the gauge parameter $\xi$. The gauge dependence is very simple, and any change in the gauge parameter can be absorbed into the coupling constant $C$ through the redefinition 
\begin{eqnarray} 
  C'=C\left(\frac34+\frac14\xi\right)\,.
\end{eqnarray}
In the Feynman gauge we have $C=C'$, whereas in the Landau gauge $C=\frac 43 C'$. We will work in the Feynman gauge, so  the mass function 
(\ref{eq:massfunc}) becomes
\begin{eqnarray} 
 M(p^2)=A(p^2)+m_0= C
 \, h^2(m^2)h^2(p^2)+m_0\,.\qquad
 \label{eq:massconstrainta}
\end{eqnarray}

The coupling constant $C$ is not a free parameter but constrained by
the mass condition (\ref{eq:massconstraint}). In the chiral limit ($m_0=0$), which we indicate by a subscript $\chi$, the constraint is
\begin{eqnarray} 
 M_\chi(m_\chi^2)=C_\chi\,h^4(m_\chi^2)
  =m_\chi\, .  \label{eq:chiralmf}
\end{eqnarray}
We choose $h(m_\chi^2)=1$, fixing the constant  at 
\begin{eqnarray} 
C_\chi=m_\chi\,.\label{eq:cl}
\end{eqnarray}
Away from the chiral limit ($m_0>0$), we expand $C$ around $C_\chi$:
\begin{eqnarray} 
 C=m_\chi+c_1 m_0+\mathcal O(m_0^2)\,
\end{eqnarray}
where $c_1$ is a constant. For sufficient small $m_0/m$ we keep the first order term only. Then the mass condition becomes
\begin{eqnarray} 
M(m^2)=(m_\chi+c_1 m_0)\,h^4(m^2) +m_0=m\, ,\label{eq:gapeq}
\end{eqnarray}
which determines the dressed mass $m$ in terms of the quark form factor $h$.
The coefficient $c_1$ will be chosen in Section~\ref{sec:massfunction} in order to give a reasonable fit to the existing LQCD data.

The final result for the dressed quark mass function is
\begin{eqnarray} 
 M(p^2)= (m_\chi+c_1 m_0)\,h^2(m^2)h^2(p^2)+m_0\,\label{eq:massfunction}
\end{eqnarray}
which reduces in the chiral limit to
\begin{eqnarray} 
 M_\chi(p^2)= m_\chi h^2(p^2)\,.\,\label{eq:chiralmf}
\end{eqnarray}
This mass function $M(p^2)$ incorporates asymptotic freedom correctly: in the ultraviolet limit the quark mass approaches the bare quark mass $m_0$, i.e.,
\begin{eqnarray} 
 \lim_{p^2\to\infty}M(p^2)
 =m_0\,.\label{eq:massfunction-asym}
\end{eqnarray}
On the other hand, in the infrared region the constituent quark mass is generated from the dressing through the interaction kernel. Even in the chiral limit of vanishing bare quark mass  a finite constituent quark mass is generated dynamically. In the present simple model for the interaction kernel where the Lorentz scalar-vector mixing in the confining interaction has been chosen such that there is no contribution to the self-energy, the dressing of the quark is entirely determined in terms of the quark form factors at the interaction vertices. 

To counter any impression that the mass function is  purely phenomenological we point out that it acquires the simple form~(\ref{eq:massfunction}) \emph{only} in the present special case of a $2:1$ Lorentz scalar-vector mixing for the confining interaction. If $\lambda\neq2$ there is a non-trivial influence from the confining interaction on the mass function, reflected in a wave function renormalization $Z\neq1$.  Furthermore, the mass function (\ref{eq:massfunction}) is the solution of the one-body CST-Dyson equation of Eq.~(\ref{eq:CSTgapE}) when the simple interaction kernels discussed in Sections~\ref{eq:decoupl} and~\ref{sec:strongff} are used.  In this model, it is the kernel that is phenomenological, with parameters fixed by data and lattice calculations as discussed below.

\section{Results and discussion}
\label{sec:massfunction}

We take a simple form for the strong quark form factors $h(p^2)$,
\begin{eqnarray}
 h(p^2)=\left(\frac{\Lambda^2_\chi-m_\chi^2}{\Lambda^2-p^2}\right)^n\,, \label{eq:hff}
\end{eqnarray}
with $\Lambda=m+M_g$ and $\Lambda_{\chi}=m_{\chi}+M_g$ where $M_g$ is an adjustable mass parameter. This form for $h(p^2)$ guarantees that there will always be one and only one mass solution of the gap equation. $n$ has to be chosen such that the pion form factor calculation described in Ref.~II converges, which requires $n>1$. In this work we choose $n=2$.

Note that $h(p^2)$ has a pole at $p^2=\Lambda^2$.  This point lies far outside of the region of 
 interest, but it may still be worthwhile to point out briefly how this result could be improved.  One possibility is to choose a function of the form
 \bea
 h(p^2)=\begin{cases}\displaystyle{\left(\frac{\Lambda^2_\chi-m_\chi^2}{\Lambda^2-p^2}\right)^2} & {\rm if}\; p^2<s_+\cr
\displaystyle{ \left(\frac{\Lambda^2_\chi-m_\chi^2}{\Lambda^2+p^2-2s_+}\right)^2} & {\rm if}\; p^2>s_+ \end{cases} \, 
 \eea
and $s_+<\Lambda^2$. This function is symmetric about $p^2=s_+$ and the pole at $p^2=\Lambda^2$ is removed. We leave further discussion of this for another time.

\begin{figure}[h!]
    \includegraphics[height=5.5cm]{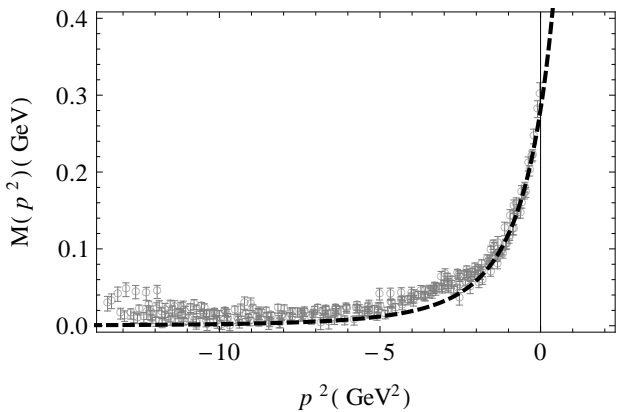} 
\caption{The quark mass function in the chiral limit fitted to the chiral-limit extrapolation of the LQCD data~\cite{Bowman:2005vx}. Data points up to $-p^2=1.94~\mathrm{GeV}^2$ have been fitted with a $\chi^2/{\rm d.o.f} = 0.61$.}\label{fig:massfunction1}
\end{figure}

The mass function in the chiral limit, Eq.~(\ref{eq:chiralmf}), involves two free parameters, the chiral constituent quark mass $m_\chi$ and $M_g$. The constituent quark mass $m$  is obtained by solving the mass condition, Eq.~(\ref{eq:gapeq}), for finite bare quark masses $m_0$. We fix our mass-function parameters using the LQCD data~\cite{Bowman:2005vx}, with four data sets for the bare quark masses $m_0=0.016, 0.032, 0.047$ and  $0.063$ GeV. Since the lattice data are calculated in Euclidean space, they must  be compared with the Minkowski-space calculation at negative $p^2$. To carry out the fit  the four sets of lattice data are extrapolated to zero bare quark mass by a linear fit, and  then  $M_g$ and $m_\chi$ are determined from a $\chi^2$ fit of the chiral mass function (\ref{eq:chiralmf}) to these (extrapolated) chiral-limit lattice data. Only the lattice data points at small values of $-p^2$  (in particular, the first 50 points between $p^2=0~\mathrm 
{GeV}^2$ and $p^2=-1.94~\mathrm 
{GeV}^2$) are used in the fit.  The reason for not using all available lattice data is that the 
ultraviolet tails of the data do not approach the corresponding values of the bare quark mass, a deviation from the correct asymptotic behavior attributed to an insufficiently small lattice spacing~\cite{Bowman:2002bm}, whereas the finite lattice spacing effects are expected to be small for the lower 50 points. This procedure gives a chiral constituent mass of $m_\chi=0.308~\mathrm{GeV}$ and a mass parameter $M_g=1.734~\mathrm{GeV}$.   
The corresponding mass function is plotted in Fig.~\ref{fig:massfunction1}. 

\begin{figure}[h!]
    \includegraphics[height=5.5cm]{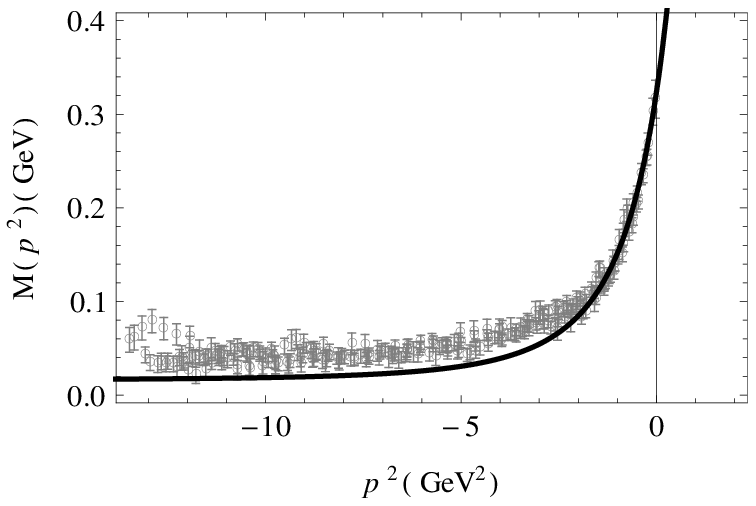} 
    \includegraphics[height=5.5cm]{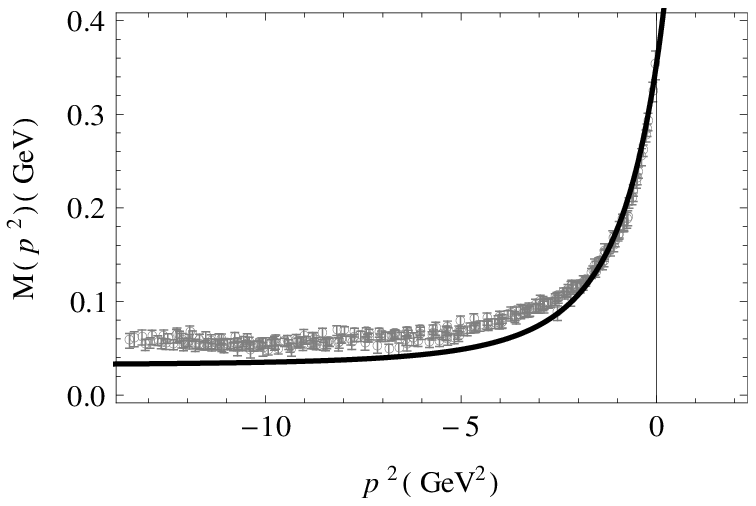} 
\caption{The quark mass function with parameters obtained from the fit to the chiral-limit extrapolation of the LQCD data~\cite{Bowman:2005vx} up to  $-p^2=1.94~\mathrm{GeV}^2$. The top figure shows the mass function for $m_0=0.016$~GeV with $m=0.363$~GeV and the corresponding lattice data, comparing with data points up to $-p^2=1.94~\mathrm{GeV}^2$ with a $\chi^2/{\rm datum} = 0.35$. The bottom figure shows the mass function for $m_0=0.032$~GeV with $m=0.403$~GeV and the corresponding lattice data, comparing with data points up to $-p^2=1.94~\mathrm{GeV}^2$ with a $\chi^2/{\rm datum} = 0.48$.}\label{fig:massfunction3}
\end{figure}

\begin{figure}[h!]
    \includegraphics[height=5.5cm]{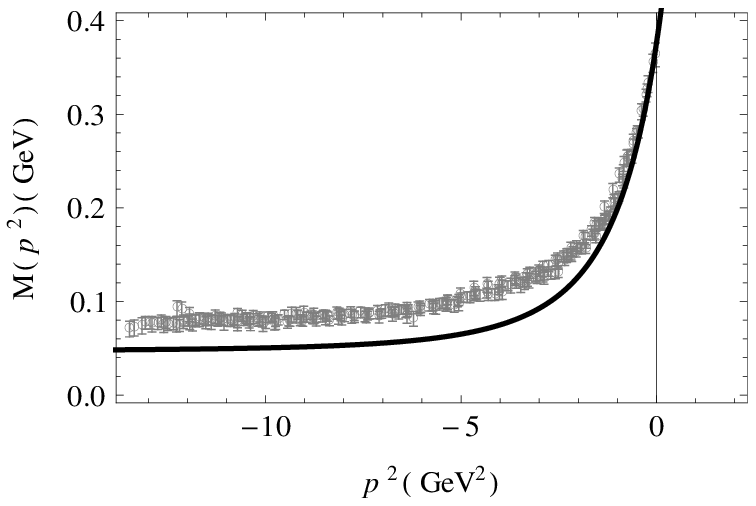} 
    \includegraphics[height=5.5cm]{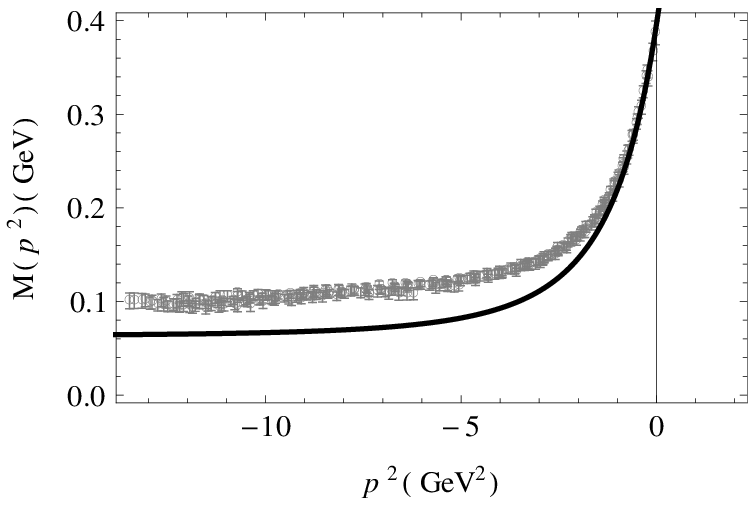} 
\caption{The quark mass function with parameters obtained from the fit to the chiral-limit extrapolation of the LQCD data~\cite{Bowman:2005vx} up to  $-p^2=1.94~\mathrm{GeV}^2$. The top figure shows the mass function for $m_0=0.047$~GeV with $m=0.434$~GeV and the corresponding lattice data, comparing with data points up to $-p^2=1.94~\mathrm{GeV}^2$ with a $\chi^2/{\rm datum} = 2.83$. The bottom figure shows the mass function for $m_0=0.063$~GeV with $m=0.462$~GeV and the corresponding lattice data, comparing with data points up to $-p^2=1.94~\mathrm{GeV}^2$ with a $\chi^2/{\rm datum} = 2.15$.}\label{fig:massfunction4}
\end{figure}

To test our parameter-fixing procedure, the mass functions for the bare quark mass values $m_0=0.016, 0.032, 0.047$ and  $0.063$ GeV have been determined by finding the dressed masses for these values of $m_0$ using the gap equation (\ref{eq:gapeq}), and comparing the results with the corresponding lattice data. With a value $c_1=12$ for the expansion coefficient in $C$ a good fit to the lattice data at small negative $p^2$ is achieved (fitting the data points up to $-p^2=1.94~\mathrm{GeV}^2$ with an overall $\chi^2/{\rm d.o.f} = 1.46$), as shown in Figs.~\ref{fig:massfunction3} and~\ref{fig:massfunction4}. 

For small $m_0$ near the chiral limit and for small $-p^2$ our results are in good agreement with the lattice data. As $m_0$ becomes larger a deviation from the lattice data is observed not only in the perturbative region of large $-p^2$ (due to finite lattice-spacing effects as mentioned above) but also at small $-p^2$. The fits using only 50 lattice data points seem to give slightly better agreement with the data for large $m_0$ than a fit of more points up to larger values of $-p^2$. In particular, it is interesting to observe in Fig.~\ref{fig:massfunction4} that for large $m_0$, the lattice data at small $-p^2$ seem to lie slightly above our results. This might be attributed to the fact that with $\lambda=2$ we have $Z=1$, which does not agree with the lattice data at small values of $-p^2$. 

Certainly, there exist various functional choices for the form factor $h(p^2)$ 
that would give a better fit to the lattice data than the simple choice~(\ref{eq:hff}). 
For instance, Ref.~\cite{Bowman:2002bm} gives a functional form suggesting that $n\sim1/2$ would be favored. However, the aim of the present work is not to provide the most accurate description of the quark mass function possible, but rather to show that our model is \emph{capable} of giving sensible results for both the quark and the pion structure\emph{ at the same time}. In Ref.~II  the pion charge form factor is calculated using the \emph{same} strong quark form factors and mass functions  obtained from the lattice data, and good agreement is obtained.

\section{Summary and conclusions}\label{sec:conclusions}

The present work is a Minkowski-space study of the dressed quark mass function using the Covariant Spectator Theory (CST).  We propose a manifestly covariant model for the interquark interaction based on previous work by GMS that incorporates both spontaneous chiral symmetry breaking and confinement. For the treatment of the light $q\bar q$ mesons such as the pion we employ a four-channel CST equation that is invariant under charge conjugation. In the nonrelativistic limit, the CST equations reduce to the Schr\"{o}dinger equation. This makes our approach suitable for a unified description of all $q\bar q$ mesons.

Spontaneous chiral-symmetry breaking is included via a Nambu--Jona-Lasinio-type mechanism: 
It is shown analytically that in the chiral limit of a vanishing current quark mass $m_0$ the two-body CST-BS equation for a pseudoscalar bound state becomes identical to the scalar part of the one-body CST-Dyson equation for the self-energy. This property ensures the existence of a zero-mass solution (a Goldstone pion) in the chiral limit. A finite dressed quark mass is then generated dynamically through the self-interactions of the quark with the $q\bar q$ interaction kernel. The present approach differs from previous CST models in the sense that the mass function is calculated directly from the kernel, which makes this model completely self-consistent.

Our interaction kernel is a covariant generalization of the nonrelativistic linear confining potential plus a constant potential shift that defines the energy scale. The confinement part of the kernel has the property that it does not contribute to the CST equation for a pseudoscalar bound state in the chiral limit. This decoupling of confinement from chiral-symmetry breaking permits our confining part to include, e.g. a Lorentz-scalar coupling as suggested from phenomenological approaches and LQCD studies. In particular, we employ a mixed  scalar-vector Lorentz structure for the confining and a pure Lorentz vector structure for the constant part of our kernel.  Furthermore, we introduce (strong) quark form factors at the interaction vertices in order to include, approximately, additional gluonic corrections and to ensure the necessary convergence in loop integrations.  

Using this kernel we have calculated the quark mass function from the CST-Dyson equation. We have chosen a particular Lorentz scalar-vector mixing in the confinement part for which the mass function is solely determined by the constant part of the kernel and by the quark form factors. Our mass function involves three free parameters which have been determined by a $\chi^2$ fit to LQCD data extrapolated to the chiral limit. For small $m_0$ and for small negative Minkowski-space quark momenta squared, $p^2$, our results are in good agreement with the lattice data. As $m_0$ becomes larger a deviation from the lattice data is observed, not only in the perturbative region, where finite-spacing effects influence the lattice results, but also at very small negative $p^2$, where the lattice results for $Z$ do not agree with our predictions. 
Applications of this work to the pion form factor is discussed in the accompanying paper, Ref.~II.

\appendix
\begin{widetext}
\section*{Appendix: Proof of charge conjugation invariance for the CST equations} \label{app:A}

Here we prove the invariance of the four coupled-channel CST equations~(\ref{eq:CST4ch}) under the substitutions (\ref{eq:C1}). 
To do this, simplify Eqs.~(\ref{eq:CST4ch}) by using the assumption that the kernel can be written as a sum of Lorentz-invariant functions that depend on momentum transfer only. With this assumption we can define six independent kernels:

\bea
&&V_d\equiv  {\cal V}_{1+,1+}={\cal V}_{1-,1-}={\cal V}_{2+,2+}={\cal V}_{2-,2-}={\cal V}(\hat p-\hat k)
\nonumber\\
&&V_s\equiv{\cal V}_{2+,2-} = {\cal V}_{2-,2+} = {\cal V}_{1+,1-} ={\cal V}_{1-,1+}={\cal V}(\hat p+\hat k) 
\nonumber\\
 &&V_{d-P}\equiv{\cal V}_{1+,2+}={\cal V}_{2-,1-} ={\cal V}(\hat p-\hat k-P) 
\nonumber\\
&&V_{s-P}\equiv{\cal V}_{1+,2-} ={\cal V}_{2-,1+} ={\cal V}(\hat p+\hat k-P) 
\nonumber\\
&&V_{d+P}\equiv{\cal V}_{2+,1+} ={\cal V}_{1-,2-}={\cal V}(\hat p-\hat k+P) 
\nonumber\\
&& V_{s+P}\equiv {\cal V}_{2+,1-}={\cal V}_{1-,2+} ={\cal V}(\hat p+\hat k+P) \, .
\eea
 With this notation the equations become
 \bea
\Gamma_{1+}( p)=-\frac12\int_k&&\Big[V_{d}\Lambda (\hat k)\Gamma_{1+}(k)S(\hat k-P)
+V_{d-P} S(\hat k+P)\Gamma_{2+}(k)\Lambda (\hat k)
\nonumber\\
&&+V_{s} \Lambda (-\hat k)\Gamma_{1-}(k)S(-\hat k-P)
+V_{s-P} S(-\hat k+P)\Gamma_{2-}( k)\Lambda (-\hat k) \Big]
\nonumber\\
\Gamma_{2+} (p)=-\frac12\int_k&&\Big[V_{d+P}\Lambda (\hat k)\Gamma_{1+}(k)S(\hat k-P)
+V_{d} S(\hat k+P)\Gamma_{2+}(k)\Lambda (\hat k)
\nonumber\\
&&+V_{s+P} \Lambda (-\hat k)\Gamma_{1-}( k)S(-\hat k-P)
+V_{s} S(-\hat k+P)\Gamma_{2-}(k)\Lambda (-\hat k)\Big]
\nonumber\\
\Gamma_{1-}(p)=-\frac12\int_k&&\Big[V_{s}\Lambda (\hat k)\Gamma_{1+}(k)S(\hat k-P)
+V_{s+P} S(\hat k+P)\Gamma_{2+}(k)\Lambda (\hat k)
\nonumber\\
&&+V_{d} \Lambda (-\hat k)\Gamma_{1-}(k)S(-\hat k-P)
+V_{d+P}S(-\hat k+P)\Gamma_{2-}(k)\Lambda (-\hat k)\Big]
\nonumber\\
\Gamma_{2-}(p)=-\frac12\int_k&&\Big[V_{s-P}\Lambda (\hat k) \Gamma_{1+}(k)S(\hat k-P)
+V_{s} S(\hat k+P)\Gamma_{2+}(k)\Lambda (\hat k)
\nonumber\\
&&+V_{d-P} \Lambda (-\hat k)\Gamma_{1-}(k)S(-\hat k-P)
+V_{d} S(-\hat k+P)\Gamma_{2-}(k)\Lambda (-\hat k)\Big] \, .
\eea
If the solution of these equations is charge conjugation invariant, then we expect the conditions (\ref{eq:C1}) to hold.
Transforming the equations for $\Gamma_{1\pm}$ (using $\eta=1$ and dropping the arguments) gives
 \bea
{\cal C}^{-1}\Gamma^T_{1+}{\cal C}=-\frac12\int_k&&\Big[V_{d}S(-\hat k+P){\cal C}^{-1}\Gamma^T_{1+}{\cal C}\Lambda (-\hat k)
+V_{d-P} \Lambda (-\hat k){\cal C}^{-1}\Gamma^T_{2+}{\cal C}S(-\hat k-P)
\nonumber\\
&&+V_{s} S(\hat k+P) {\cal C}^{-1}\Gamma^T_{1-}{\cal C}\Lambda (\hat k)
+V_{s-P} \Lambda (\hat k) {\cal C}^{-1}\Gamma^T_{2-}{\cal C}S(\hat k-P)\Big]
\nonumber\\
\to \Gamma_{2-}=-\frac12\int_k&&\Big[V_{d}S(-\hat k+P)\Gamma_{2-}\Lambda (-\hat k)
+V_{d-P} \Lambda (-\hat k)\Gamma_{1-}S(-\hat k-P)
\nonumber\\
&&+V_{s} S(\hat k+P)\Gamma_{2+}\Lambda (\hat k)
+V_{s-P}\Lambda (\hat k) \Gamma_{1+}S(\hat k-P)\Big]  
\nonumber\\
{\cal C}^{-1}\Gamma^T_{2+}{\cal C}=-\frac12\int_k&&\Big[V_{d+P} S(-\hat k+P) {\cal C}^{-1}\Gamma^T_{1+}{\cal C}\Lambda (-\hat k)
+V_{d} \Lambda (-\hat k) {\cal C}^{-1}\Gamma^T_{2+}{\cal C} S(-\hat k-P)
\nonumber\\
&&+V_{s+P} S(\hat k+P) {\cal C}^{-1}\Gamma^T_{1-}{\cal C} \Lambda (\hat k)
+V_{s} \Lambda (\hat k) {\cal C}^{-1}\Gamma^T_{2-}{\cal C}S(\hat k-P)\Big]
\nonumber\\
\to\Gamma_{1-}=-\frac12\int_k&&\Big[V_{d+P} S(-\hat k+P) \Gamma_{2-}\Lambda (-\hat k)
+V_{d} \Lambda (-\hat k) \Gamma_{1-} S(-\hat k-P)
\nonumber\\
&&+V_{s+P} S(\hat k+P) \Gamma_{2+}\Lambda (\hat k)
+V_{s} \Lambda (\hat k) \Gamma_{1+}S(\hat k-P)\Big]\,.
 \eea
This shows that the transformations (\ref{eq:C1}) transform the $\Gamma_1$ equations into the $\Gamma_2$ equations, proving they are equal.  Similarly, the $\Gamma_2$ equations can be transformed into the $\Gamma_1$ equations, completing the argument.

\end{widetext}
\begin{acknowledgements}
E.B. and F.G. are pleased to thank Craig D. Roberts for valuable discussions and for providing the lattice data sets. This work received financial support from Funda\c c\~ao para a Ci\^encia e a Tecnologia (FCT) under Grants No.~PTDC/FIS/113940/2009, No.~CFTP-FCT (PEst-OE/FIS/U/0777/2013) and No.~POCTI/ISFL/2/275. This work was also partially supported by the European Union under the HadronPhysics3 Grant No. 283286, and by Jefferson Science Associates, LLC under U.S. DOE Contract No. DE-AC05-06OR23177.
\end{acknowledgements}

\bibliographystyle{h-physrev3}
\bibliography{PapersDB-v1.3E}

\end{document}